\documentclass[astronomy,article,accept,pdftex,moreauthors]{Definitions/mdpi} 

\firstpage{1} 
\pubvolume{1}
\issuenum{1}
\articlenumber{0}
\pubyear{2025}
\copyrightyear{2025}
\externaleditor{Firstname Lastname} 
\datereceived{ } 
\daterevised{ } 
\dateaccepted{ } 
\datepublished{ } 
\hreflink{https://\\doi.org/} 

\usepackage{graphicx}
\newcommand{\vsini}{\ensuremath{{\upsilon}\sin i}}

\newcommand{\Teff}{$T_\mathrm{eff}$}
\newcommand{\logg}{$\log g$}

\Title{The Status of the Astrophysical Parameters of Upper Main Sequence Stars}

\TitleCitation{The Status of the Astrophysical Parameters of Upper Main Sequence Stars}

\Author{Lukas Kueß $^{1}$\orcidA{} and Ernst Paunzen $^{2,}$\highlighting{*}
\orcidB{}}
 
\AuthorNames{Lukas Kueß and Ernst Paunzen}
\isAPAStyle{%
       \AuthorCitation{Kueß L., Paunzen E.}
         }{%
        \isChicagoStyle{%
        \AuthorCitation{Lukas Kueß, Ernst Paunzen}
        }{
        \AuthorCitation{Kueß, L.; Paunzen, E.}
        }
}

\address{%
$^{1}$ \quad  Department of Astrophysics, Vienna University, Türkenschanzstraße 17, A-1180 Vienna, Austria\\
$^{2}$ \quad  Department of Theoretical Physics and Astrophysics, Faculty of Science, Masaryk University, Kotl\'{a}\v{r}sk\'{a} 2, \mbox{Brno 611 37, Czechia;} {a01405976@unet.univie.ac.at} 
}
\corres{epaunzen@physics.muni.cz}

\abstract{Calibrating the ages, masses, and radii of stars on the upper main sequence depends heavily on accurate measurements of the effective temperature (\Teff) and surface gravity (\logg). These parameters are difficult to obtain meticulously due to the nature of hot stars, which exhibit features such as rapid rotation, atomic diffusion, pulsation, and stellar winds. We compare the \Teff\, and \logg\, values of apparent normal B-F stars in four recent catalogues that employ different methods and pipelines to obtain these parameters. We derived various statistical parameters to compare the differences between the catalogues and discussed the astrophysical implications of these differences. Our results show that the huge differences in \Teff\, (up to $10^4$\,K) and \logg\, (up to 2 dex) between the catalogues have serious implications on the determination of ages, masses, and radii of the stars in question. We conclude that there appears to be no homogeneous set of stellar parameters on the upper main sequence, and one must be cautious when interpreting results obtained from using only one of the catalogues. The homogenisation of said parameters is an essential task for the future and will have a significant impact on astrophysical research dealing with stars on the upper main sequence.}

\begin{document}

\section{Introduction}

The upper main sequence of stars is populated 
by core hydrogen-burning objects with masses larger than approximately 1.4\,M$_{\odot}$. Their stellar
atmospheres are typically characterised by no, or only thin, convection layers, with the flux transported bradiation~\citep{2022ApJS..262...19J}. These thin convection layers of helium and iron are essential for driving pulsations in these stars \citep{1999AcA....49..119P}.
The disappearance of general convection in stellar atmospheres is believed to happen around
6800\,K, or a spectral type of about F2\,V \citep{2002ApJ...564L..93D}.

With this abrupt offset of strong convection in the atmospheres and envelopes, the overall
rotational velocities change, and processes like diffusion, meridional circulation, and mass loss
become important \citep{1991ApJ...370..693C}. Furthermore, strong (up to tens of kG) stable
magnetic fields complicate the proper description of the stellar atmospheres \citep{2001ASPC..248..305M}.
With objects of the upper main sequence, several important astrophysical processes can be studied
in more detail. The $\delta$ Scuti variables show pulsation driven mainly through the opacity 
$\kappa$ mechanism operating in the helium-ionising zone. For the cooler pulsating stars, coupling 
between convection and pulsation is critical. It affords an explanation for the $\gamma$ Doradus 
variables located at the red border of the classical pulsation strip
\citep{2019MNRAS.490.4040A}. The hotter $\beta$ Cephei
variables (spectral types between O9 and B4) are driven by the opacity mechanism operating in the 
partial ionisation zone of iron-like metals. The exact mechanism is responsible for the Slowly Pulsating 
B-type (SPB) stars. In these less massive stars, the iron opacity bump is located in a deeper layer 
\citep{2015A&A...580L...9W}. Another essential aspect of upper main sequence stars is chemical
peculiarity. 

They were identified by \citet{1897AnHar..28....1M}. 
Most initial scientific inquiry was concerned with identifying 
distinctive attributes within their spectra and thoroughly assessing their photometric patterns.
The ingredients for this phenomenon are chemical separation, diffusion effects, atmospheric mixing,
stellar winds, and accretion
in the (non-) presence of an organised stellar magnetic field \citep{2023MNRAS.519.5913A,2025A&A...694A.270U}.

For many applications in astronomical research, it is important to have accurate estimations of astrophysical parameters. Among these parameters are two crucial values that determine the properties of a star and its location in the Hertzsprung--Russell diagram (HRD): (i) the effective temperature, \Teff\, in [K], and (ii) the surface gravity, \logg\, in dex, derived from $g$ in [cm s$^{-2}$].

Let us recall that surface gravity is not independent of the effective temperature but can be expressed as
\begin{equation*}
\log g = \log M + 4\log T_\mathrm{eff} - \log L   
\end{equation*}
including the stellar mass ($M$) and luminosity ($L$). Notice that there exists a dependence 
of rotation on the mass and luminosity \citep{1990ApJS...74..501P,2002ApJ...573..359A}.

In the literature, many calibrations for lower main sequence stars exist 
\citep{2007A&A...475..519H,2016A&A...595A..15T}. In contrast, upper main sequence stars have mainly been
calibrated using the Strömgren--Crawford and Geneva photometric systems \citep{2005A&A...444..941P,2006A&A...458..293P}. The reasons are manifold, mainly because consistent
models considering all the astrophysical processes described above are unavailable.
Furthermore, as the peak of the blackbody radiation curve shifts more to the blue and ultraviolet region
(the Balmer jump is at its maximum at 10,000\,K), the effect of interstellar 
reddening, i.e., the UV bump \citep{1989ApJ...345..245C} becomes significant. 
This is important for calculating absolute fluxes and fitting the spectral energy distribution \citep{2024RASTI...3...89M}.

Another important aspect is the metallicity and its influence on the stellar formation
and evolution \citep{2001A&A...371..152M}. The composition of a star changes its effective temperature and luminosity for the same mass and age. The spread of metallicity in the solar neighbourhood \citep{2004A&A...420..183A} and among members of open clusters \citep{2010A&A...523A..71G} is within a factor of ten on the main sequence. Other aspects are different stellar populations \citep{2024A&A...690A.107S} and the metallicity gradient within the Milky Way due to radial elemental migration
\citep{2022MNRAS.509..421N}. 

In this paper, we assess the current status of the basic astrophysical parameters \Teff\ and \logg\
taken from four available and widely used corresponding catalogues. For this, we defined a
sample of ``normal'' upper main-sequence stars based on spectra from the Large Sky Area Multi-Object Fiber Spectroscopic Telescope \citep[LAMOST,][]{2012RAA....12..723Z}. Then, we investigated the 
differences between the objects in common, searched for offsets, and derived statistical 
parameters of the distributions. To our knowledge, this is the first such attempt, and we could not find any mention in the literature that discusses the discrepancies of stellar parameters in different catalogues.

We have not investigated the metallicities included in the four references. This is caused by
the very different definitions of metallicity, such as [Fe/H], [Z], and [M/H]. It is simply
impossible to transform these different values into one standard system. In fact, this is
an essential task for the future.

We discuss the possible effects of measured projected rotational velocity (\vsini) and compare some astrophysical relations dependent on these parameters.

\section{Target Selection and Catalogues} \label{tsac}

We started our target selection with stars from the so-called \textit{Golden OBA sample} by 
\citet{2023A&A...674A..39G}, which were observed within the LAMOST survey. 
Furthermore, we included stars with temperatures higher than 6000\,K from the
LAMOST LRS DR9 Stellar Parameter Catalog of A, F, G, and K Stars.

The LAMOST DR9 low-resolution spectra have a spectral resolution of R\,$\sim$1800 and 
cover the wavelength range from 3700--9000\,\AA. We used only spectra with a signal-to-noise
ratio (S/N) larger than 50 in $g$. This lower boundary was chosen based on the fact that spectra with lower S/N can result in misclassification in the following step.
Spectral classifications were derived with the modified
MKCLASS code {(MKCLASS} 
 is a computer program in C used 
to classify stellar spectra on the Morgan--Keenan--Kellman (MKK) spectral classification 
scheme \citep{2014AJ....147...80G}. The installation guide and further information can 
be accessed {at} 
 \url{https://www.appstate.edu/\~grayro/mkclass/} URL accessed on 27 10 2024)
{of} 
 \citet{2020A&A...640A..40H}, specifically adapted to detect chemically peculiar stars. Discarding all stars with a peculiar classification and luminosity classes between III and I
(non-main-sequence objects), we obtained a sample of 63,660 objects.

The following four sources of astrophysical parameters were used for our analysis. They include widely 
different data sets and algorithms for their calculations.

{\it \citet[]{2019AandA...628A..94A,2022AandA...658A..91A} [StarHorse2021]}{: They} 
 combined parallaxes 
and photometry from the $Gaia$ DR3 with the photometric catalogues
of Pan-STARRS 1 \citep{2013ApJS..205...20M}, 2MASS \citep{2006AJ....131.1163S}, AllWISE \citep{2013yCat.2328....0C}, 
and SkyMapper DR2 (\citet[]{2019PASA...36...33O}) [without the $u$ filter] to derive
Bayesian stellar parameters, distances, and extinctions. For this purpose, the StarHorse code \citep{2018MNRAS.476.2556Q} was
applied. This is a Bayesian parameter estimation code that compares many observed quantities to stellar evolutionary
models. Given the set of observations plus several priors, it finds the posterior probability over a
grid of stellar models, distances, and extinctions. It was concluded that the systematic
errors of the astrophysical parameters are smaller than the nominal uncertainties for most objects. We refer to this catalogue as ``StarHorse'' or ``SH21'' in the text and figures.

{\it \citet[]{2023AandA...674A..28F} [Gaia DR3 Apsis]}: This is the pipeline software (Astrophysical Parameter Inference System, Apsis) developed by the $Gaia$ consortium.
They analysed astrometry, photometry, BP/RP, and RVS spectra for objects across the Hertzsprung--Russell diagram. 
Their method was compared and validated with star cluster data, asteroseismological results, and several other
references.
It is claimed that despite some limitations, this is the most extensive catalogue of uniformly inferred stellar parameters to date.

{\it \citet[]{2022MNRAS.514.4588L} [LASSO-MLPNet]}: They used a neural network-based scheme to deliver atmospheric parameters from low-resolution stellar spectra of LAMOST.
They used a polynomial fitting method to estimate and remove the pseudo-continuum for every spectrum. Secondly, they detected parameter-sensitive features using the Least Absolute Shrinkage and Selection Operator (LASSO). This evaluates the effectiveness of spectral 
features based on the combined effects of spectral fluxes and the noise overlapped on them. Finally, they proposed a multilayer perceptron neural network-based method (MLPNet) to 
estimate the stellar atmospheric parameters.
There is a restriction on the \Teff\ range to 13,000\,K, or a spectral type of B8\,V. 

{\it \citet[]{2019AJ....158..138S} [The Revised TESS Input Catalog]}: This procedure is based on the 
apparent magnitude in the $TESS$ bandpass ($T$), taking into account the stellar evolutionary phases
(dwarfs, subgiants, and giants divided by their \logg\ values).
In their analysis, they used PHOENIX model atmospheres \citep{2013A&A...553A...6H}, combining photometric and spectroscopic data with Gaia DR2 and 2MASS calibrations. All calibrations are listed in \citet{2018AJ....156..102S}.
For our analysis, we used v8.2 of the catalogue. This catalogue is called ``TIC'' in our work.

The rotational velocities were taken from two catalogues. The first is from \cite{2024A&A...689A.141S}, who took spectra from the LAMOST DR9 medium-resolution survey (MRS) and inferred the parameters \Teff, \logg, \vsini, and $[M/H]$ via the Stellar LAbel Machine (SLAM, \cite{2020ApJS..246....9Z}), which was trained on ATLAS12 \citep{2005MSAIS...8...25C} atmospheres beforehand.

The second catalogue that includes rotational velocities is from \cite{2024ApJS..271....4Z}. Here, the authors used spectra from the low-resolution survey (LRS) and MRS and from LAMOST DR9. The rotational velocities were derived by fitting synthetic PHOENIX spectra \citep{2013A&A...553A...6H} to the observed ones. The method was validated on the Apache Point Observatory Galactic Evolution Experiment \citep{2015mwss.confE..47M}.

\begin{figure}[H]
    \includegraphics[width=\textwidth]{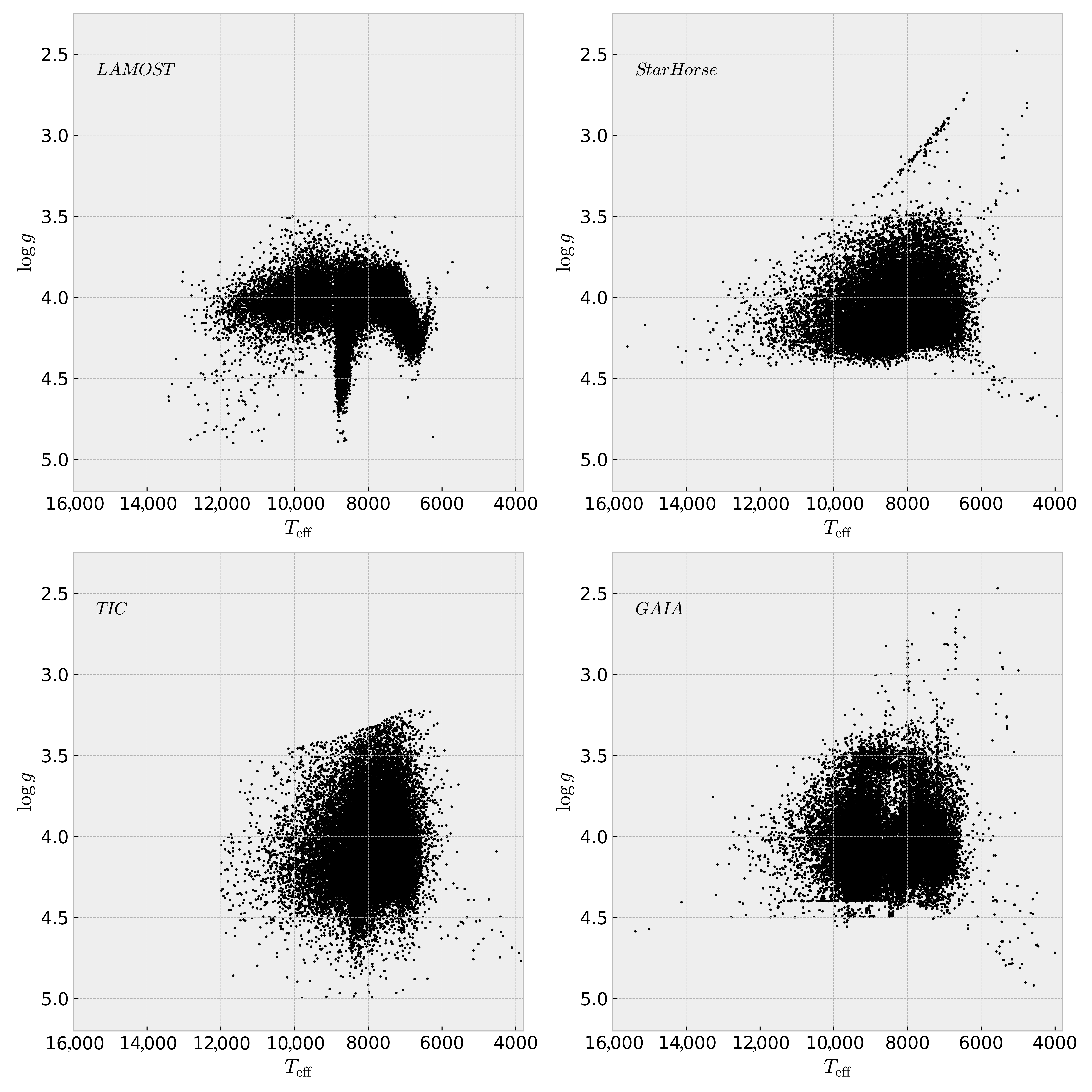}
    \caption{{Kiel} 
 diagrams for the different catalogues we used.}
    \label{fig:hrd}
\end{figure}

In Figure~\ref{fig:hrd}, we plot the Kiel diagram for 31,503 stars that are listed in each of the four catalogues in our analysis. It is easy to see that there are significant inhomogeneities between the samples.

\section{Methods}

\subsection{Kinematic Selection}

\begin{figure}[H]
    \includegraphics[width=\textwidth]{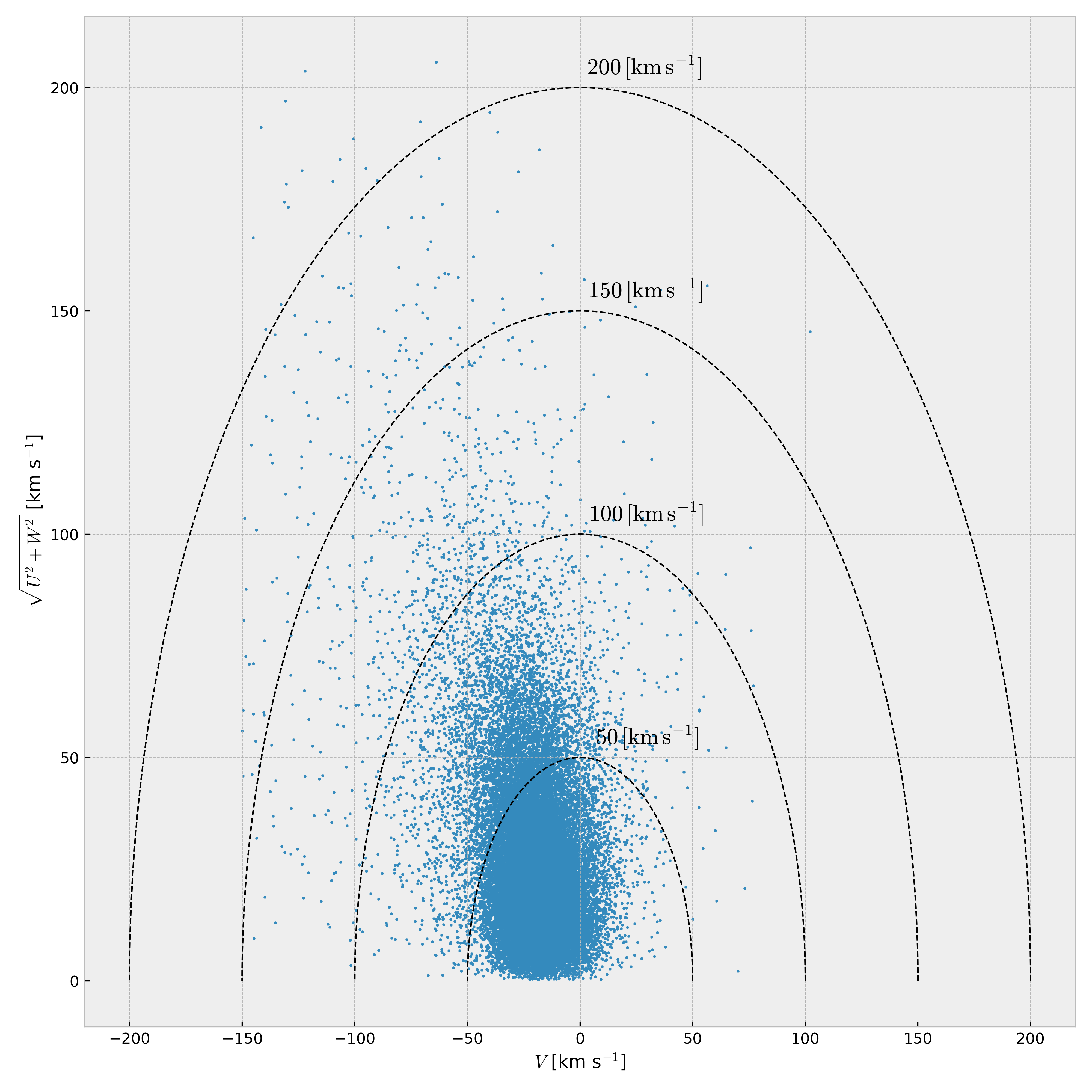}
    \caption{Toomre diagram of the stars in our sample. The dotted lines describe total velocities $V_{tot}$ in increments of 50 [$\mathrm{km\,s}^{-1}$].}
    \label{fig:toomre}
\end{figure}

To discriminate between the stars of Population I (galactic disk) and II (galactic halo) in the list of stars that are mentioned in all catalogues, we made some cuts in the spatial velocities $U$, $V$, and $W$. For this, we utilised a Toomre diagram. We used the criterion of \citet{2014A&A...562A..71B} to identify the populations, which included a cut at $V_{tot} = \sqrt{U_{LSR}^2 + V_{LSR}^2 + W_{LSR}^2}$ = 200 $\mathrm{km\,s}^{-1}$. Stars with $V_{tot}<200\,\mathrm{km\,s}^{-1}$ are considered to be in the galactic (thin and thick) disk, while stars with $V_{tot}>200\,\mathrm{km\,s}^{-1}$ are likely halo stars and thus belong to Pop II.
Figure \ref{fig:toomre} shows the Toomre diagram with lines of equal total velocity $V_{tot}$. Note that there are very few stars outside the boundary of $V_{tot} = 200\,\mathrm{km\,s}^{-1}$. Keeping them in the sample would not lead to differences in our results in a significant way.

\subsection{Extinction Measurements} \label{extinction}

A common practice is using photometric calibrations to derive astrophysical parameters
\citep{1982VilOB..60...16S,2006MNRAS.371.1793K}. For this, an estimation of the reddening (extinction, 
or absorption) mainly caused by interstellar dust is needed. 
An error in determining the 
reddening will have severe consequences, for example, for drawing colour magnitude diagrams and 
determining astrophysical parameters, such as the effective temperature, luminosity, mass, and age~\citep{2006MNRAS.371.1641P,2009A&A...501..941H}. Most of the references discussed in this paper also
present reddening estimates in different wavelength bands. However, these estimates must also be
treated with caution. We analysed the listed values as a supplement to the published astrophysical
parameters.

\begin{figure}[H]
    \includegraphics[width=\textwidth]{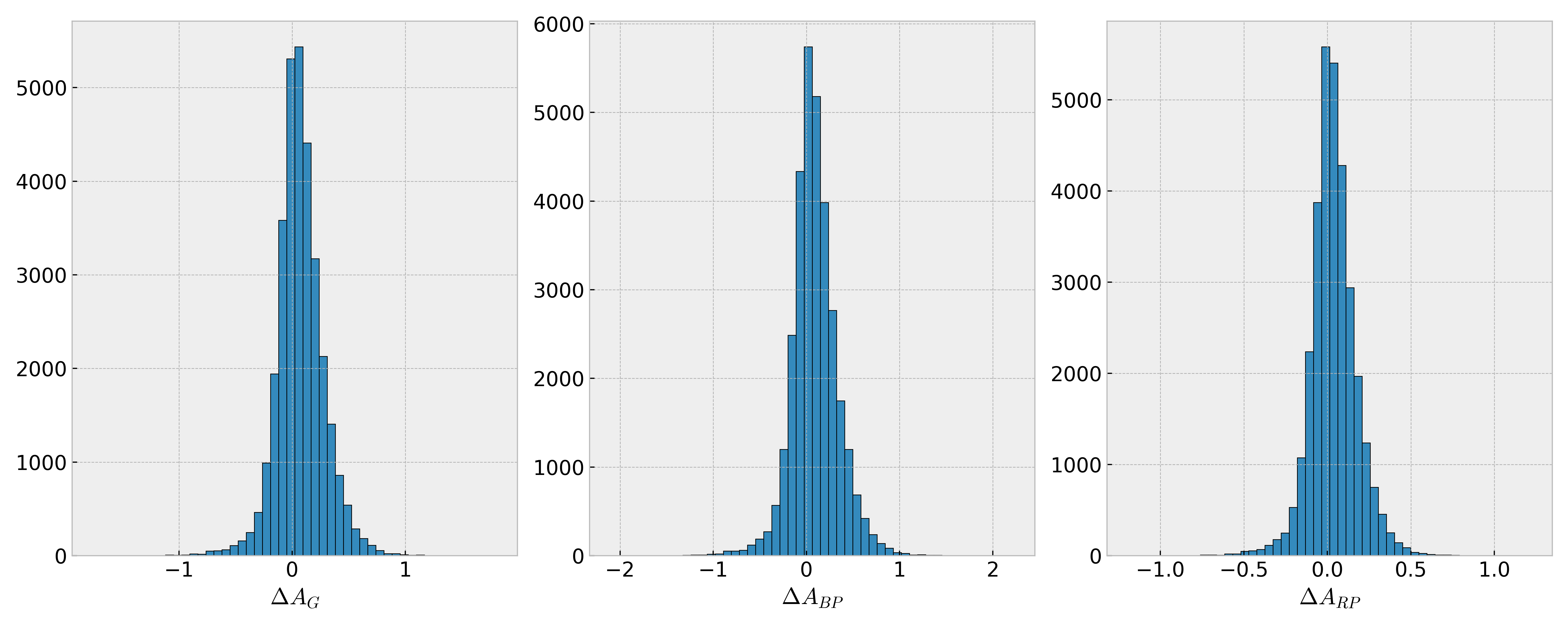}
    \caption{Comparison between the extinction values from $Gaia$ and StarHorse. The x-axes are defined via $\Delta X = X_{Gaia} - X_{SH}$.}
    \label{fig:ext_hist}
\end{figure}

As an example, we compare the $Gaia$ DR3 extinction measurements ($A_G$, $A_{BP}$, $A_{RP}$) to the ones derived by $StarHorse2021$. A histogram of the differences can be seen in Figure~\ref{fig:ext_hist}.

In the end, we also calculated a few key statistical properties of the distribution:

 \begin{itemize}
     \item Mean ($\mu$);
     \item Standard deviation ($\sigma$);
     \item Skewness ($\gamma_1$);
     \item Kurtosis ($\gamma_2$).
 \end{itemize}

Skewness measures, if a dataset is symmetric with respect to its mean. A commonly used criterion is that a distribution is considered symmetric if that parameter lies in the interval $\gamma_1\in [-0.5, 0.5]$.  Kurtosis measures the shape of the tails of the distribution or how much a specific dataset is spread around the mean.

\begin{table}[H]
 \caption{{Statistical} 
 properties of the differences in extinction.}
    \label{tab:ext_stats}
    \setlength{\tabcolsep}{21.5pt}
    \begin{tabular}{ccccc}
    \midrule
    \textbf{Difference} & \boldmath{$\mu$} \textbf{[Mag]} & \boldmath{$\sigma$} \textbf{[Mag]} & \boldmath{$\gamma_1$} & \boldmath{$\gamma_2$}\\ 
    \midrule
       $\Delta A_G$ & +0.072 & +0.211 & -0.013 & +4.111 \\
       $\Delta A_{BP}$ & +0.091 & +0.251 & +0.034 & +4.248 \\
       $\Delta A_{RP}$ & +0.037 & +0.140 & -0.026 & +4.297 \\
       \midrule
    \end{tabular}
\end{table}

\begin{figure}[H]
    \includegraphics[width=\textwidth]{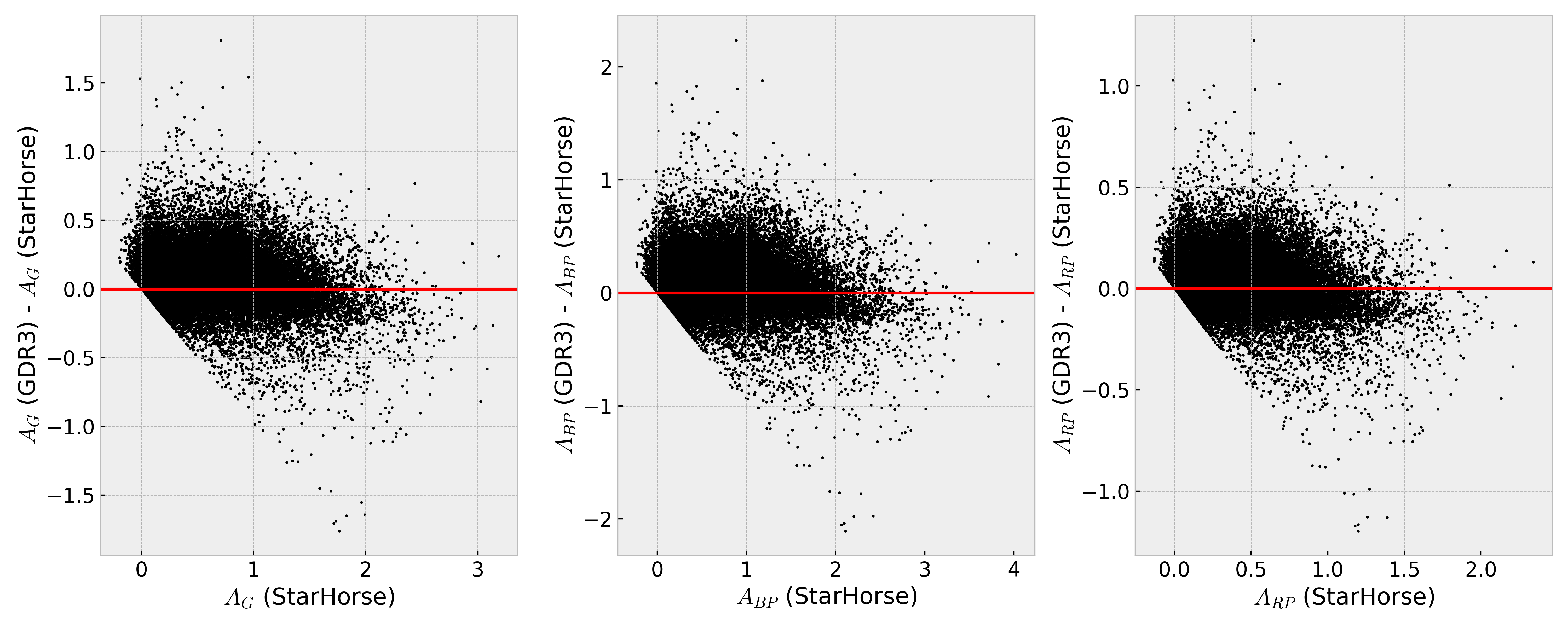}
    \caption{{Scatter} 
 plot version of Figure~\ref{fig:ext_hist}. A clear cut of unknown reason can be seen. The horizontal red lines show the ideal case of the two catalogues having the same value.}
    \label{fig:ext_scatter}
\end{figure}

Figure~\ref{fig:ext_scatter} shows the differences as a function of the extinction measurements from $StarHorse2021$ {and Table~\ref{tab:ext_stats} shows some statistical properties of these distributions}. There is a clear-cut trend in the scatter plot. From the references, 
we were not able to deduce the reason for it. However, it shows that one must be very cautious 
using the published reddening values of individual objects.

\vspace*{-12pt}

\section{Results} \label{results}

\begin{figure}[H]
    \includegraphics[width=0.9\textwidth]{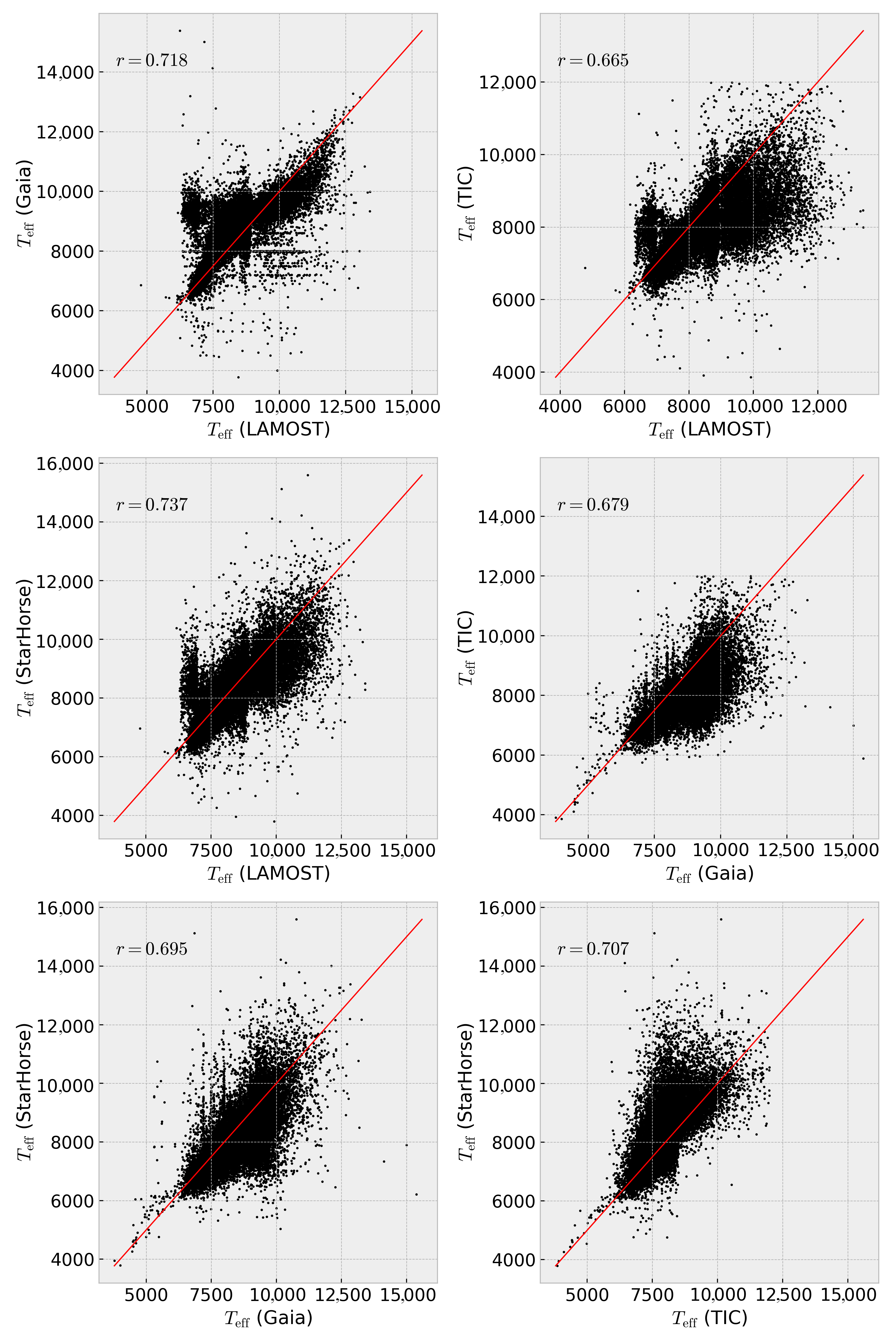}
    \caption{{The} 
 correlation plots and r-values for \Teff\,(Section \ref{tsac}). The red line in each plot shows a 1:1 relation.}
    \label{fig:teff_dist}
\end{figure}

\begin{figure}[H]
    \includegraphics[width=\textwidth]{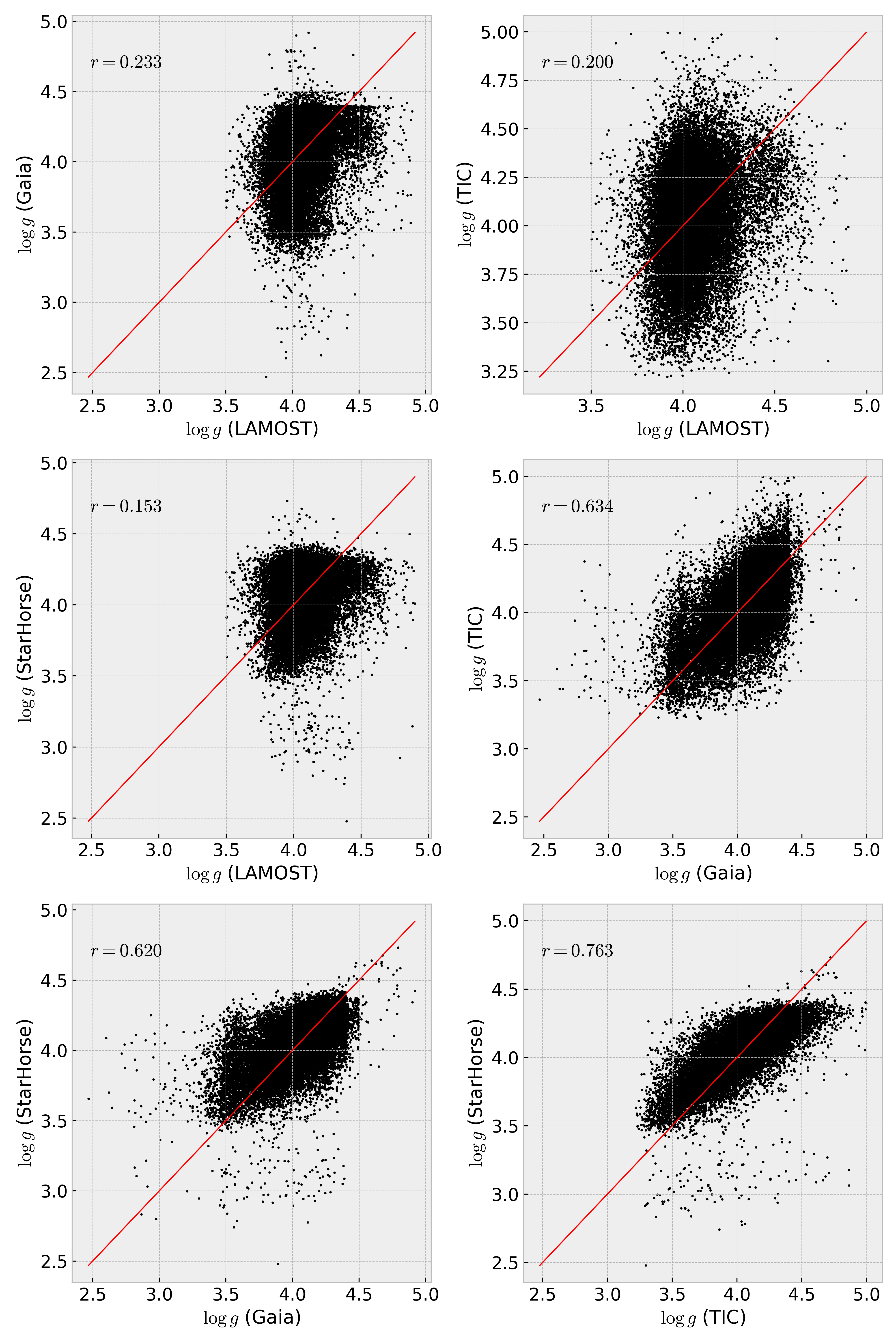}
    \caption{Same as Figure~\ref{fig:teff_dist} but for $\log g$.}
    \label{fig:logg_dist}
\end{figure}

 In total, 31,503 stars have data for \Teff\, and \logg\, in all four catalogues. This is about half of our initial sample. This number results in cross-matching losses and using only the stars that have both, \Teff\, and \logg\, values listed. We compared the parameters by simply plotting them as a function of each other. The resulting diagrams can be seen in Figure~\ref{fig:teff_dist} for the effective temperature and in Figure \ref{fig:logg_dist} for the surface gravity. The red line in each subplot shows a 1:1 relationship, and the r-value in the plot denotes the Pearson coefficient of the relation, given by

\begin{equation}
    r = \frac{cov(X,Y)}{\sigma_X\sigma_Y}
\end{equation} 

 Additionally, we plotted the Kiel diagrams (HRDs) for these 31\,503 stars in each of the catalogues (Figure~\ref{fig:hrd}).

\subsection{Effective Temperature}

As one can readily see (Figure~\ref{fig:teff_hist}), there is a considerable discrepancy between the \Teff\ values derived in each catalogue, up to multiple thousands, in a few cases up to $\Delta T_\mathrm{eff} = 10^4$\,K.

One can see in Table \ref{tab:teff_stats} that the mean values of the distributions, which ideally would be close to zero, are shifted by up to $\sim$460\,K, and the standard deviations are large (570--720\,K). This cannot be explained by errors in the data and the estimation of \Teff\, but may be due to systematic deviations in the different catalogues and their respective methods. It can also be seen that all but two comparisons (LAMOST--Gaia and SH21--TIC) are skewed to one side. This and the relatively large 
also suggest the possibility of outliers that influence the shape of the distributions. Some of these outliers can be seen in Figure~\ref{fig:teff_dist}.

\begin{table}[H]
\caption{Statistical properties of the differences in \Teff.}
    \label{tab:teff_stats}
    
      \setlength{\tabcolsep}{18.5pt}
    \begin{tabular}{ccccc}
    \midrule
    \textbf{Difference} & \boldmath{$\mu$} \textbf{[K]} & \boldmath{$\sigma$} \textbf{[K]} & \boldmath{$\gamma_1$} & \boldmath{$\gamma_2$}\\ 
    \midrule
       LAMOST---SH21 & +55.404 & 703.794 & +0.505 & +4.959 \\
       LAMOST---TIC & +249.421 & 721.973 & +1.204 & +4.724 \\
       LAMOST---Gaia & $-$205.857 & 708.814 & +0.014 & +8.502 \\
       SH21---TIC & +194.017 & 574.952 & +1.432 & +7.976 \\
       SH21---Gaia & $-$261.262 & 700.037 & $-$0.101 & +5.162 \\
       TIC---Gaia & $-$455.278 & 662.693 & $-$0.695 & +3.810 \\
       \midrule
    \end{tabular}
    
\end{table}

\subsection{Surface Gravity}

A similar picture is given in the distributions of $\Delta \log g$. The differences are as large, as $\Delta \log g = 2.0$\,dex, meaning a factor of 100 in the difference of the surface gravity (Figure~\ref{fig:logg_dist}). The statistical analysis reveals that three distributions (SH21--TIC, SH21--Gaia, and TIC--Gaia) are symmetric {Table~\ref{tab:logg_stats}}.

\begin{table}[H]
   \caption{{Statistical} 
 properties of the differences in \logg.}
    \label{tab:logg_stats}
       \setlength{\tabcolsep}{19.5pt}
    \begin{tabular}{ccccc}
    \midrule
   
    \textbf{Difference} & \boldmath{$\mu$} \textbf{[dex] }& \boldmath{$\sigma$} \textbf{[dex]} & \boldmath{$\gamma_1$} & \boldmath{$\gamma_2$}\\ 
    \midrule

       LAMOST---SH21 & $-$0.009  & +0.211 & +0.829 & +1.724 \\
       LAMOST---TIC & $-$0.012 & +0.251 & +0.616 & +0.435 \\
       LAMOST---Gaia & +0.022 & +0.217 & +0.885 & +1.652 \\
       SH21---TIC & $-$0.002 & +0.125 & +0.233 & +2.923 \\
       SH21---Gaia & +0.032 & +0.138 & +0.013 & +4.137 \\
       TIC---Gaia & +0.034 & +0.165 & $-$0.146 & +2.056 \\
       \midrule
    \end{tabular}
 
\end{table}

Here, the mean values agree more, but the standard deviation is also relatively large at values up to 0.251\,dex. This is almost a factor two in the difference of dex in \logg. Again, this cannot simply be explained by errors in data and numerical errors in processing them.

\subsection{Correlation with \vsini}

Additionally, we wanted to see whether there is a dependence on \vsini\, in the differences of \Teff\ and \logg. 
We plotted \vsini\ as a function of differences in the different catalogues. In Figures \ref{fig:teff_fang} and \ref{fig:logg_fang}, one can see the plots for the catalogue of \cite{2024ApJS..271....4Z}, and in Figures \ref{fig:teff_sun} and \ref{fig:logg_sun}, one can see the ones for \cite{2024A&A...689A.141S}. No trend is visible in any of those plots, which is not too surprising, as we assume that \vsini\, would have more of an effect on the star's metallicity than on the other parameters since the rotation of a star can have a non-negligible effect on the convection zones in the upper layers and thus also changes the observed chemical composition. This is also made clear by the Pearson coefficient printed in each of the plots.

\subsection{Bolometric Correction} \label{BC}

To derive accurate bolometric magnitudes and thus luminosities, one needs to correct the absolute $V$-magnitude of a star via a bolometric correction, $BC$:

\begin{equation}
    M_{Bol} = M_V + BC_V
\end{equation}

\begin{figure}[H]
   \includegraphics[width=\textwidth]{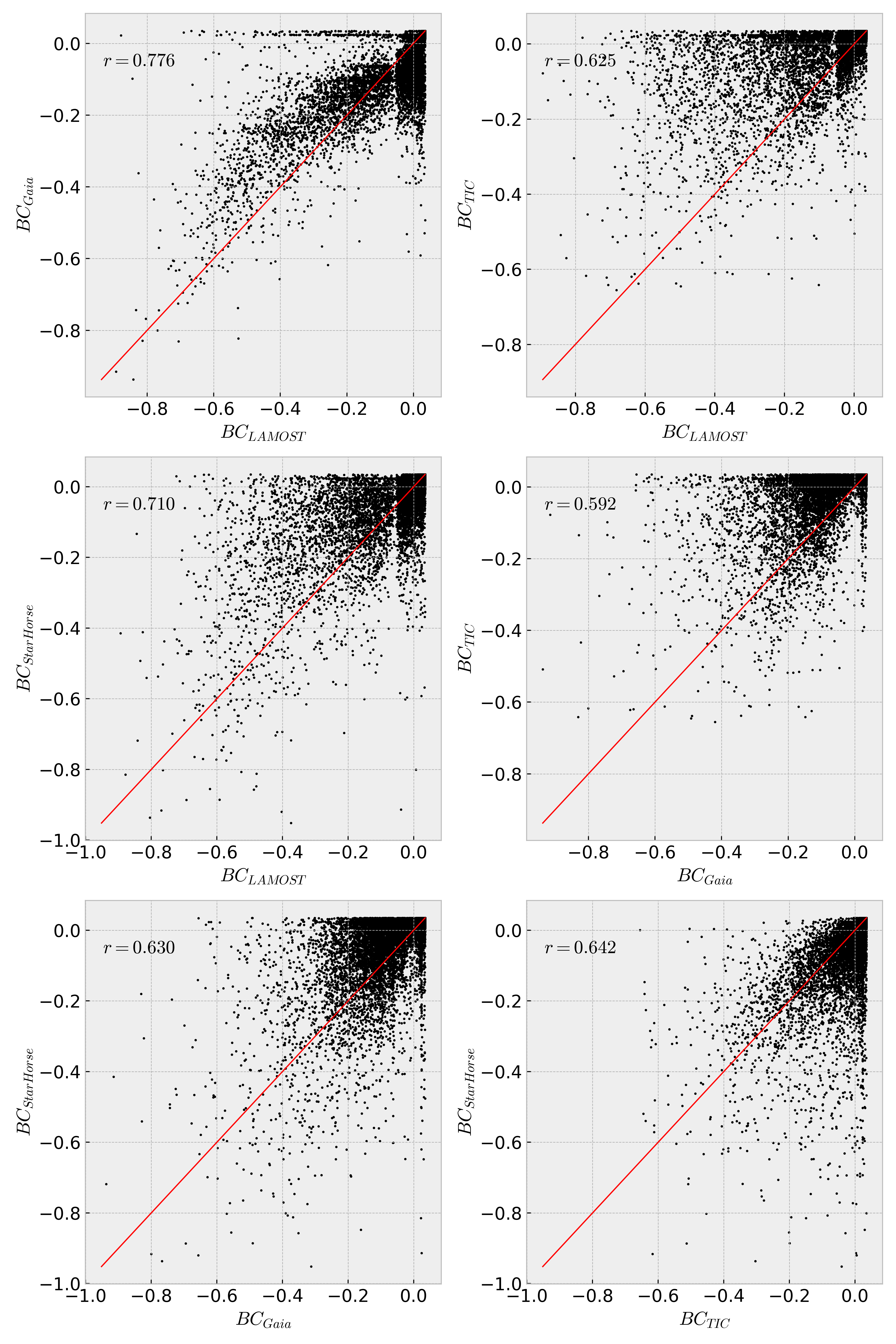}
    \caption{{Comparison} 
 of $BC_V$ values from the different catalogues. {The red lines correspond to a 1:1 relationship.}}
    \label{fig:bc_comp}
\end{figure}

This $BC_V$ is a polynomial function of \Teff. In this work, we use the polynomials from \cite{1996ApJ...469..355F} with their correction from \cite{2010AJ....140.1158T}. Equation \eqref{eq:BC} shows this polynomial function, and the coefficients can be seen in Table 1 of \citet{2010AJ....140.1158T}. 
A comparison of $BC_V$ values derived from each catalogue is seen in Figure~\ref{fig:bc_comp}.

\begin{equation}
    BC_V = a + b(\log T_\mathrm{eff}) + c (\log T_\mathrm{eff})^2 + d (\log T_\mathrm{eff})^3 + \cdots
    \label{eq:BC}
\end{equation}

Since $BC_V$ is related to the luminosity of a star via

\begin{equation}
    \frac{L_\star}{\mathrm{L}_\odot} = 10^{-0.4(M_{bol,\star}-M_{bol,\odot})}
\end{equation}
one can derive a luminosity ratio between two sources in terms of their difference in $BC_V$:

\begin{equation}
    \frac{L_1}{L_2} = 10^{-0.4 \Delta BC_V}
    \label{eq:lum_ratio}
\end{equation}

\begin{figure}[H]
 \hspace*{-6PT}
  \includegraphics[width=0.8\textwidth]{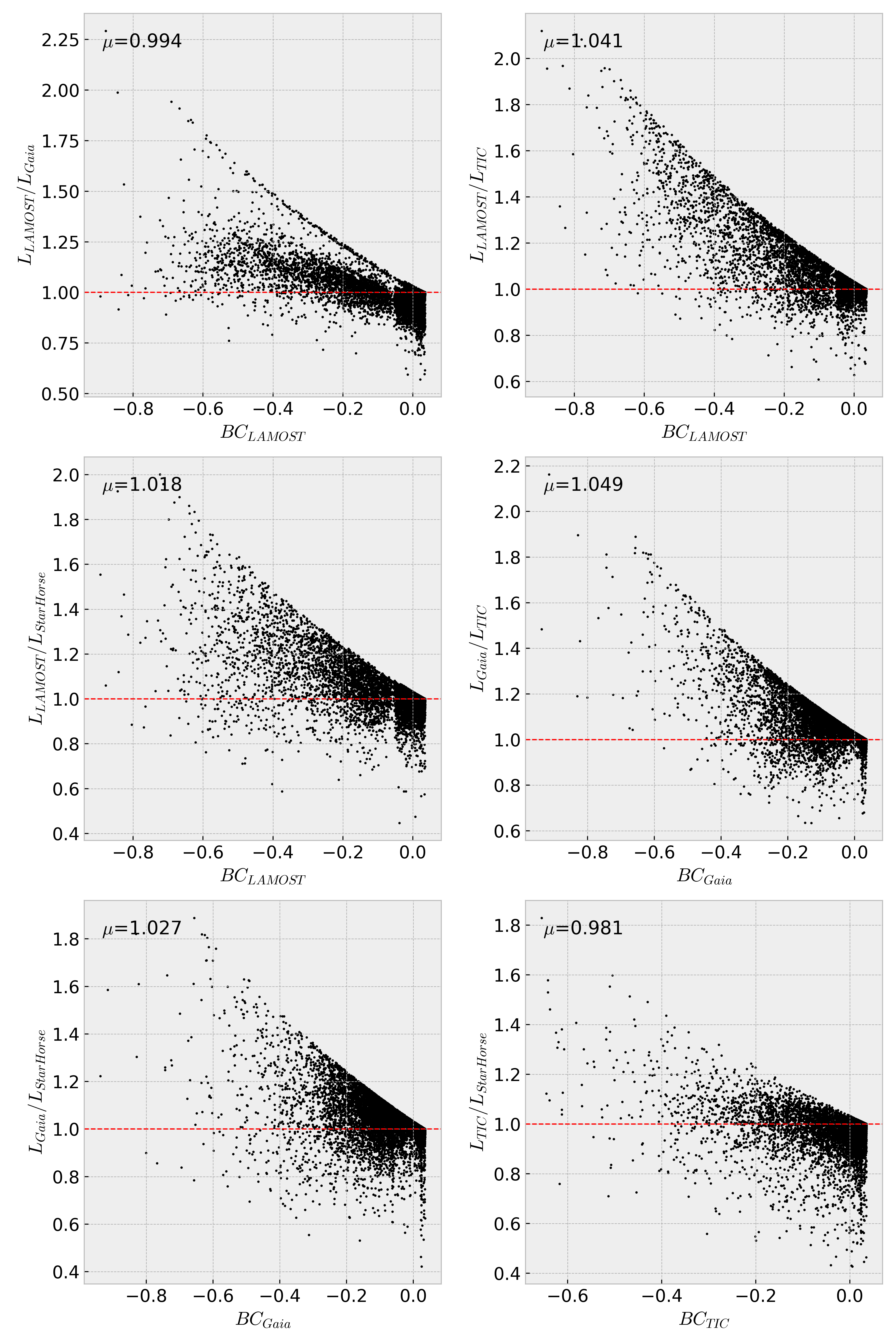}
    \caption{Same as Figure~\ref{fig:bc_comp}, but for the derived luminosity ratios. $L_1$ denotes the bolometric luminosity derived from \Teff\ in the respective catalogue; $L_2$ is the bolometric luminosity derived from the subtracted $BC_V$. The dashed red horizontal line corresponds to an ideal luminosity ratio of one.}
    \label{fig:bc_lum}
\end{figure}

This difference, as seen in Equation \eqref{eq:lum_ratio}, has implications on the calculated bolometric luminosities. As seen in Figure~\ref{fig:bc_lum}, stars of the same temperature in one catalogue can have a luminosity that differs by a factor greater than two when comparing the bolometric luminosities derived from two different catalogues of \Teff, assuming that the mass of the star stays the same.

Since the luminosity also gives an estimation of the main-sequence lifetime,

\begin{equation}
    \tau_{MS} \propto \frac{M_\star}{L_\star}
\end{equation}

This discrepancy also has implications for estimating the evolutionary status of a star. 

\subsection{Mass-Luminosity Relation}

The mass and luminosity of a star are classically connected by a mathematical relation of the form of

\begin{equation}
    L_\star \propto M_\star^\alpha
    \label{eq:mlr}
\end{equation}
 i.e., a star's luminosity is a function of its mass. The exponent $\alpha$ in this relation is dependent on the mass range one investigates. Different divisions in mass have been used in the literature, and thus different values for $\alpha$ have been derived. Reference \cite{2018A&A...619L...1W}, for example, lists three publications with distinct subdivisions in mass (see Table~\ref{tab:mass_range}).

 \begin{table}[H]

   \caption{Different slopes for the MLR found in the literature.}
     \label{tab:mass_range}

\setlength{\tabcolsep}{38pt}
     \begin{tabular}{ccc}
  
     \midrule
         \textbf{Mass Range} \textbf{[}\boldmath{$M/\mathrm{M}_\odot$}\textbf{]} & \boldmath{$\alpha$} & \textbf{Reference} \\
         \midrule
       
         $M>0.40$ & 2.440 & \cite{1988JRASC..82....1G}\\
         $0.40 < M < 5.01$ & 4.160 & \\
         $M > 5.01$ & 3.510 & \\
         \midrule
         $0.7 < M $ & 2.62 & \cite{1991ApSS.181..313D}\\
         $ M > 0.7 $ & 3.92 & \\
         \midrule
         $0.38 < M \leq 1.05$ & 4.841 & \cite{2015AJ....149..131E} \\
         $1.05 < M \leq 2.40$ & 4.328 & \\
         $2.40 < M \leq 7.00$ & 3.962 & \\
         $M > 7$ & 2.726 & \\
         \midrule
     \end{tabular}
    
 \end{table}

 In addition to a relation simply between mass and luminosity, \cite{2018ApJS..237...21M} also derive empirical relations that incorporate other parameters, such as \Teff, \logg, [Fe/H] and stellar density $\rho$. In favour of simplicity, we stick with a relation like Equation~\eqref{eq:mlr}. 
 We calculated the masses (and radii) of our stars via the fitting of isochrones using the \textit{Stellar Isochrone Fitting Tool} (StIFT ({\url{https://github.com/Johaney-s/StIFT}}), which is based on \citet{2010MNRAS.401..695M}, and used the PARSEC isochrones \citep{2012MNRAS.427..127B} to infer stellar parameters. From these radii and the \Teff\ values, the luminosities of the stars were calculated using the well-known Stefan--Boltzmann law:

 \begin{equation}
     L_\star \propto R_\star^2 T_\mathrm{eff, \star}^4
     \label{eq: stefan_boltzmann}
 \end{equation}
 
\begin{figure}[H]
   \includegraphics[width = \textwidth]{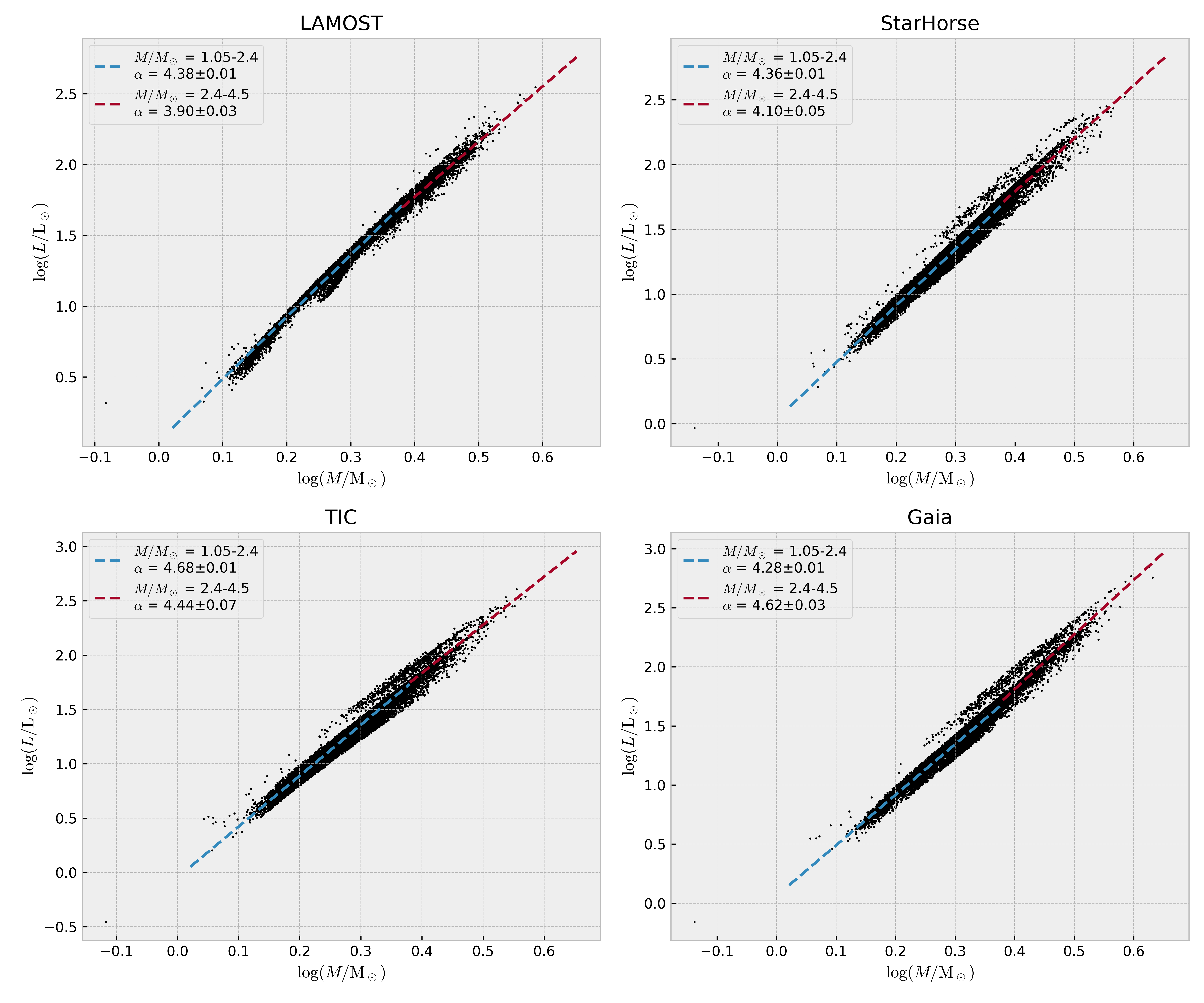}
     \caption{MLR of our sample stars.  The slopes for the piecewise functions are given in the plot and also in Table~\ref{tab:mlr}.}
     \label{fig:mlr}
 \end{figure}

 Also, we split our mass range in two, as almost all of our stars lie in the interval [1.05, 4.5]$\mathrm{M}_\odot$, which are the two ``middle'' sections in the relation by \cite{2015AJ....149..131E}. We then fit a classical MLR in these two regions. The result can be seen in Figure~\ref{fig:mlr} and in Table~\ref{tab:mlr}. One can see that the slopes are steeper than the reported values in the literature (Table~\ref{tab:mass_range}). Another intriguing characteristic is the different ``branches'' clearly 
 deviating from the line of equality. Those can be found except for the LAMOST data set. We checked 
 all possible correlations with \vsini, the age, and other astrophysical parameters. We did not find
 any satisfying explanation for this behaviour.

\begin{table}[H]
      \caption{The exponent in the MLR for each catalogue and mass range derived from them.}
    \label{tab:mlr}

    \setlength{\tabcolsep}{28pt}
    \begin{tabular}{ccc}
    \midrule
   
      \textbf{Survey}   & \shortstack{\boldmath{$\alpha$}\\ \textbf{for}\\ \boldmath{$1.05 < M/\mathrm{M}_\odot \leq 2.4$}} & \shortstack{\boldmath{$\alpha$}\\ \textbf{for}\\\boldmath{$2.4 < M/\mathrm{M}_\odot \leq 4.5$}} \\
      \midrule
     
       LAMOST & $4.38\pm0.01$ & $3.90\pm0.03$\\
       StarHorse & $4.36\pm0.01$ & $4.10\pm0.05$\\
       TIC & $4.68\pm0.01$ & $4.44\pm0.07$\\
       Gaia & $4.28\pm0.01$ & $4.62\pm0.03$\\
       \midrule
    \end{tabular}
  
\end{table}

\vspace*{-6pt}

\subsection{Mass-Radius Relation}

Similarly, we examined the relationship between stellar mass and radius. A mass--radius relation (MRR) can be derived from an MLR and a mass--temperature relation (MTR) using Equation~\eqref{eq: stefan_boltzmann} \citep{2018MNRAS.479.5491E}. The result is an equation like

\begin{equation}
    \log(R_\star) = \alpha (\log M_\star)^2 + \beta \log M_\star + \gamma
    \label{eq:mrr}
\end{equation},
which we derived in the way mentioned above.

\begin{figure}[H]
    \includegraphics[width=\textwidth]{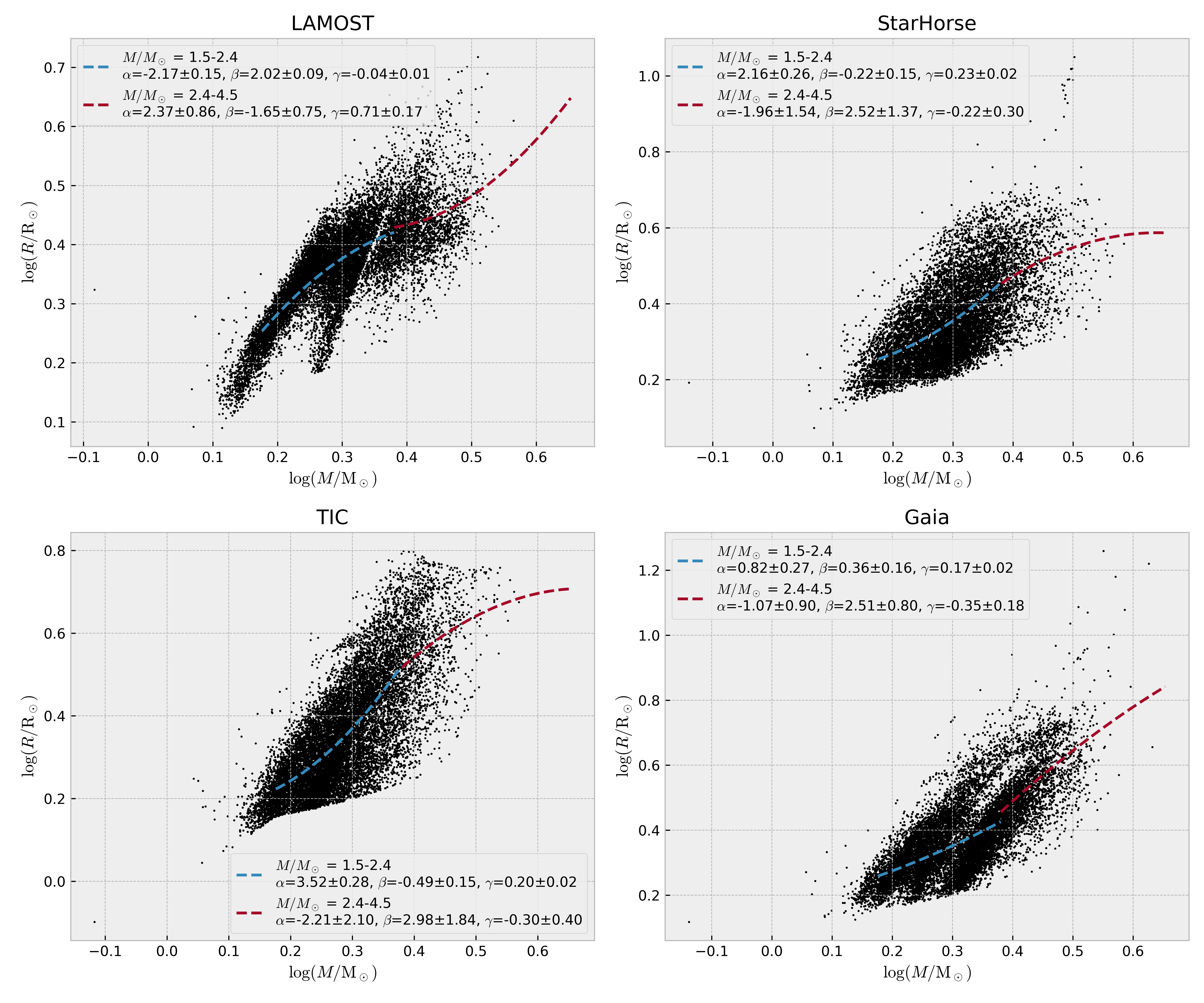}
    \caption{Mass--radius relation for our catalogues. The dashed lines correspond to the quadratics in the lower and higher-mass regimes, respectively. Again, the coefficients of the piecewise functions are given in the plots and in Table~\ref{tab:mrr}.}
    \label{fig:mrr}
\end{figure}

We again used a piecewise fit for the same mass ranges as for the MLR.  The results are seen in Figure~\ref{fig:mrr} and in Table~\ref{tab:mrr}.

\begin{table}[H]
    \caption{Coefficients for the piecewise quadratic fits in Equation~\eqref{eq:mrr}.}
    \label{tab:mrr}
    
    \setlength{\tabcolsep}{53pt}
    \begin{tabular}{cc}
        \midrule
     
        \textbf{Survey} & \textbf{Mass–Radius Relation (MRR)} \\
        \midrule
     
        \multirow{9}{*}{LAMOST}
            & \textbf{{Lower mass} 
}\\
            & $\alpha = -2.17\pm 0.15$\\
            & $\beta = +2.02\pm0.09$\\
            & $\gamma = 0.04\pm0.01$ \\
            &  \\
            & \textbf{Higher mass}\\
            & $\alpha = +2.37\pm0.86$\\
            & $\beta = - 1.65\pm0.75$\\
            & $\gamma =  + 0.71\pm0.17$ \\
        \midrule
        \multirow{9}{*}{StarHorse} 
            & \textbf{Lower mass}\\
            & $\alpha = +2.16\pm0.26$\\
            & $\beta =  -0.22\pm0.15$\\
            & $\gamma = +0.23\pm0.02$\\
            &  \\
            & \textbf{Higher mass}\\
            & $\alpha = -1.96\pm1.541$\\
            & $\beta = + 2.52\pm1.37$\\
            & $\gamma =  - 0.22\pm0.30$ \\

 \bottomrule
\end{tabular}
\end{table}

\begin{table}[H]\ContinuedFloat
\caption{{\em Cont.}}
       \label{tab:mrr}
    
    \setlength{\tabcolsep}{53pt}
    \begin{tabular}{cc}
        \midrule
     
        \textbf{Survey} & \textbf{Mass–Radius Relation (MRR)} \\
        \midrule

        \multirow{9}{*}{TIC} 
            & \textbf{Lower mass}\\
            & $\alpha = - 3.52\pm0.28$\\
            & $\beta =  - 0.49\pm0.15$\\
            & $\gamma =  0.20\pm0.02$ \\
            &  \\
            & \textbf{Higher mass}\\
            & $\alpha = -2.21\pm2.10$\\
            & $\beta =  + 2.98\pm1.84$\\
            & $\gamma =  - 0.30\pm0.40$ \\
        \midrule
        \multirow{9}{*}{Gaia} 
            & \textbf{Lower mass}\\
            & $\alpha = +0.82\pm0.27$\\
            & $\beta = + 0.36\pm0.16$\\
            & $\gamma = + 0.17\pm0.02$ \\
            & \\
            & \textbf{Higher mass}\\
            & $\alpha = -1.07\pm0.90$\\
            & $\beta = + 2.51\pm0.80$\\
              & $\gamma =  - 0.35\pm0.18$ \\
        \midrule
    \end{tabular}
   
\end{table}

\subsection{Ages of Star Clusters}

Of course, different values of \Teff\, give different luminosities. To highlight this, we matched our sample of stars to the catalogue of open clusters from \cite{2023A&A...673A.114H} to see which ones are members of clusters. We found 807 stars in 370 clusters. The cluster with the most members in our sample was NGC 2168, which contained 38 stars. To see the age differences across different catalogues, we plotted the luminosities derived from Equation \eqref{eq: stefan_boltzmann} against the temperatures in each catalogue.

\begin{figure}[H]
   \includegraphics[width=\textwidth]{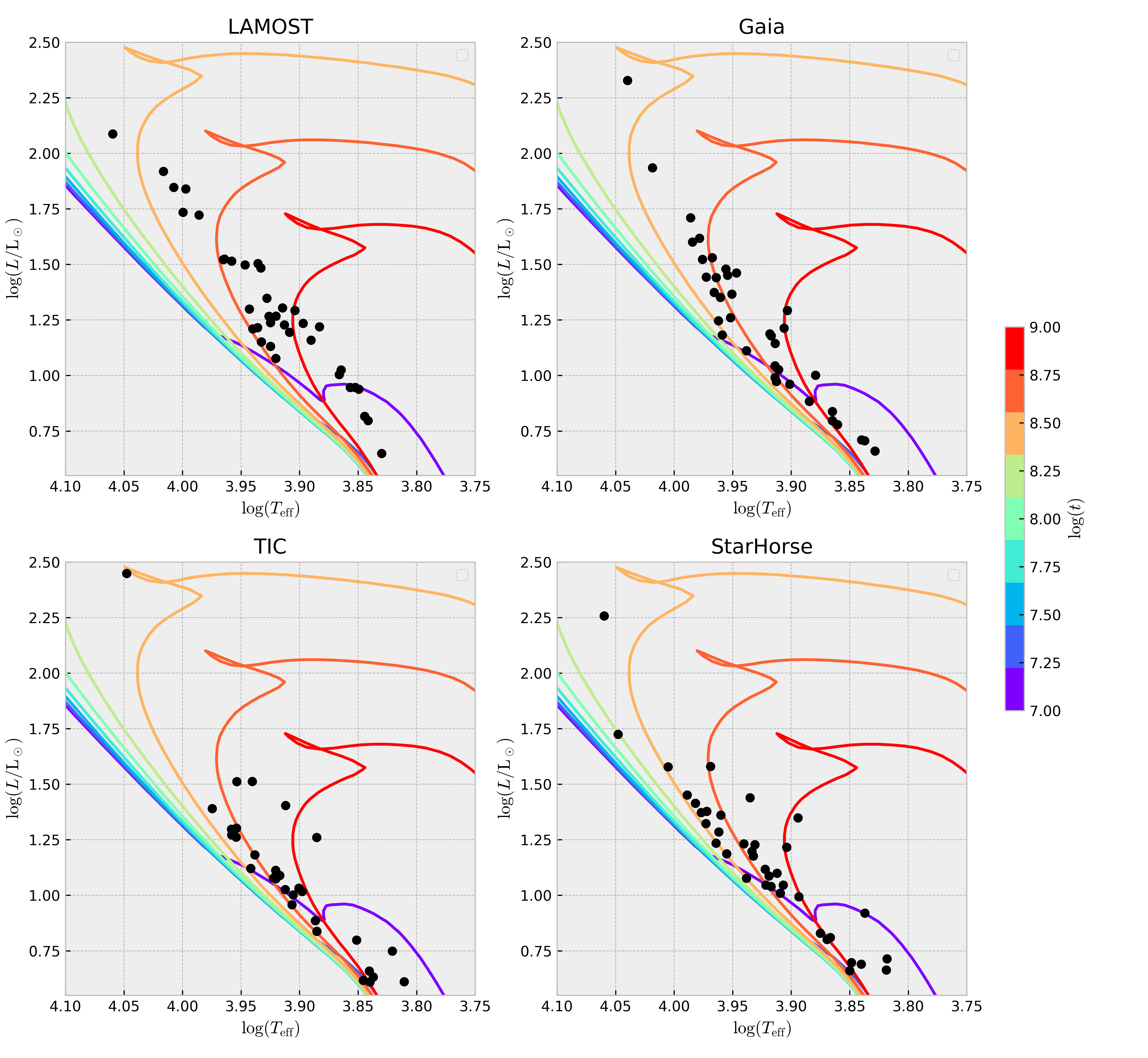}
    \caption{Different ages for the different catalogues of 38 stars in the cluster NGC 2168. The isochrones were taken from \cite{2012MNRAS.427..127B} using $Z = 0.012$.}
    \label{fig:ngc_2168}
\end{figure}

As seen in Figure~\ref{fig:ngc_2168}, there is a significant difference when determining the age of star clusters based on luminosities derived from different \Teff-values. We also plotted a range of isochrones from \cite{2012MNRAS.427..127B} between $\log \in [7.0, 9.0]$ and using $Z=0.012$ (\citet{2022MNRAS.509..421N}). The age of the cluster was estimated to be $\log t = 8.21$ \cite{2023A&A...673A.114H} or $\log t = 7.95$ (\citet{2022MNRAS.509..421N}). Figure~\ref{fig:ngc_2168} suggests that the best-fitting isochrones are either younger (TIC, StarHorse) or older ($Gaia$, LAMOST) than that value. Of course, the sample is just a small portion of the cluster, and everything comes with uncertainties. Still, one can see that the age differences vary between different sets of astrophysical parameters.

\section{Conclusions} \label{conclusions}

We compared astrophysical parameters of ``normal'' stars on the upper main sequence 
with spectral types derived from LAMOST spectra using MKCLASS. The investigated parameters 
are the effective temperature (\Teff) and surface gravity (\logg) of four 
large surveys, each with its own unique dataset and methodology, to see if they are comparable in their estimations. 
Our results show that the estimations of the parameters in the different catalogues are 
off by up to almost $\Delta T_\mathrm{eff} = 10^4$K in temperature and two dex 
in surface gravity. This has severe implications for interpreting astrophysical 
results with values taken from either of these catalogues, i.e., one cannot assume that 
one obtains similar results when working with another set of astrophysical parameters. Another danger is the determination of a rough spectral type based solely on the effective temperature, which may lead to incorrect assessments of stellar properties, such as convection and diffusion. From our analysis, we can see that the \Teff values from LAMOST and SH21 have the best correlation ($r = 0.737$), whereas the correlation between the values of LAMOST and TIC is the weakest ($r = 0.665$). For \logg, the correlations are generally weaker, resulting in the best correlation between TIC and SH21 ($r = 0.763$) and the weakest between LAMOST and SH21 ($r = 0.153$). The reasons for these weak correlations may lie in outliers that we did not remove for our analysis. Additionally, the values can differ based on the nature of the stars. A fraction of our sample stars may have been misclassified in one or more of the four catalogues, in the \textit{Golden OBA} sample, or in our spectral classification, which ultimately would, at least in part, explain the discrepancies we found.

We also analysed various astrophysical relations for each catalogue and compared them to one another. These included a relation between mass and luminosity (MLR), mass and radius (MRR), and the effects of bolometric corrections for different temperatures. We conclude that there are significant implications for main-sequence lifetime (up to a factor of two) and age estimations when compared with each other.

In conclusion, there is a need to achieve a homogeneous calibration of the mentioned stellar parameters to overcome these issues.

As mentioned in the introduction, we did not compare the metallicity values given in the 
catalogues because they all use different definitions to measure this value, making them almost impossible to compare directly.
This, too, is an essential task for future efforts.

To strengthen the conclusion we arrived at, we also plotted the HRD for each catalogue 
(Figure~\ref{fig:hrd}); one can also see that the values derived by the different 
methods in the catalogues are all over the place, and one cannot simply compare them. 

\vspace*{+6pt}

\authorcontributions{{Conceptualization, E.P., methodology: L.K and E.P., formal analysis, L.K., writing--original draft preparation: L.K. and E.P., writing--review and editing, L.K. and E.P.; visualization, L.K. All authors have read and agreed to the published version of the manuscript.}} 

\funding{{
This work was supported by the grant GA{\v C}R 23-07605S.}} 

\conflictsofinterest{{The authors declare no conflict of interest.}} 

\appendixtitles{yes} 

\appendixstart
\appendix

\section{Histograms of the Differences in \boldmath{\Teff}\, and \logg}

\begin{figure}[H]
    \includegraphics[width=\textwidth]{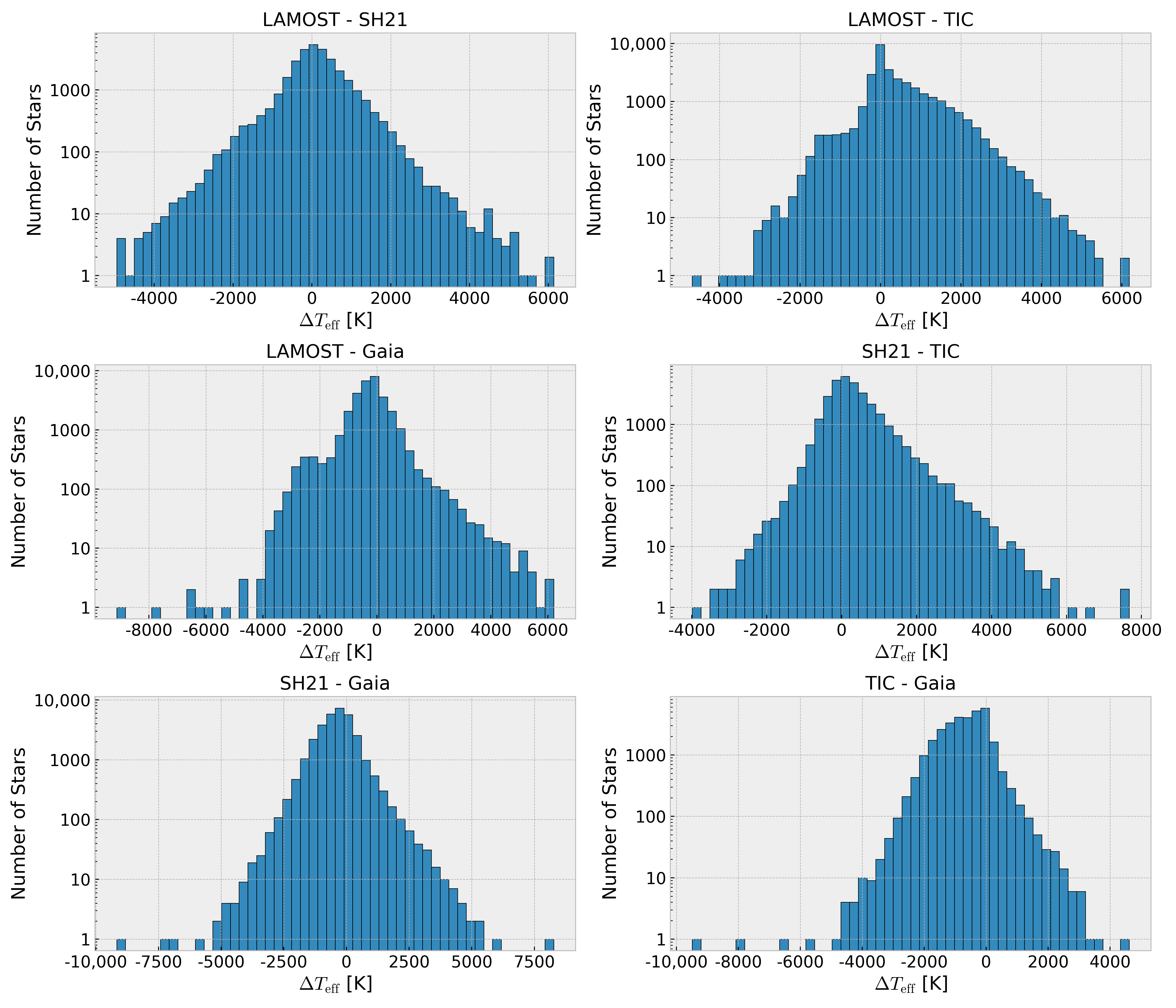}
    \caption{{Histogram} 
 of the differences in \Teff.}
    \label{fig:teff_hist}
\end{figure}

\begin{figure}[H]
    \includegraphics[width=\textwidth]{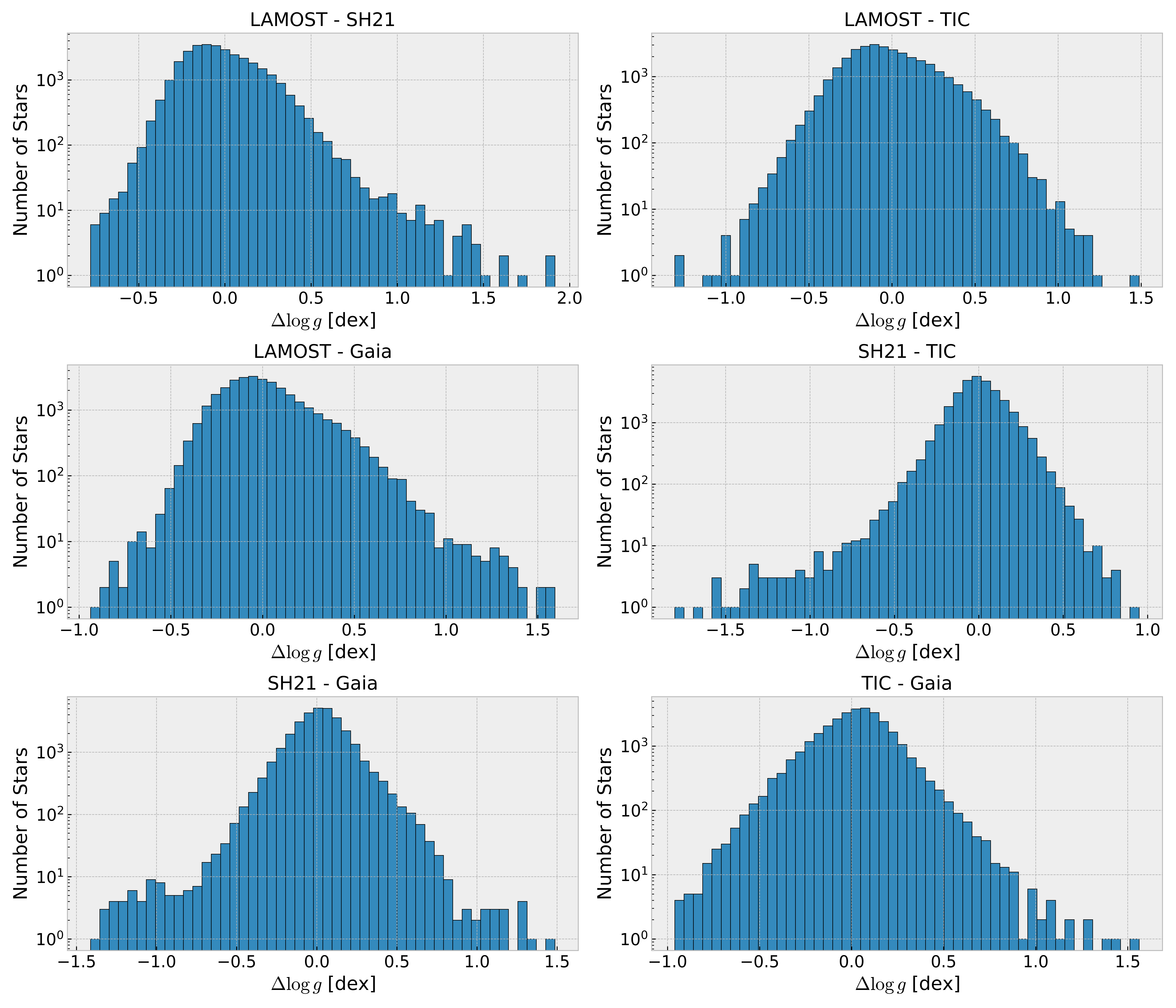}
    \caption{{Same} 
 as Figure~\ref{fig:logg_hist}, but for \logg.}
    \label{fig:logg_hist}
\end{figure}


\section{Plots of the Correlation with \boldmath{\vsini}}

\begin{figure}[H]
    \includegraphics[width=\textwidth]{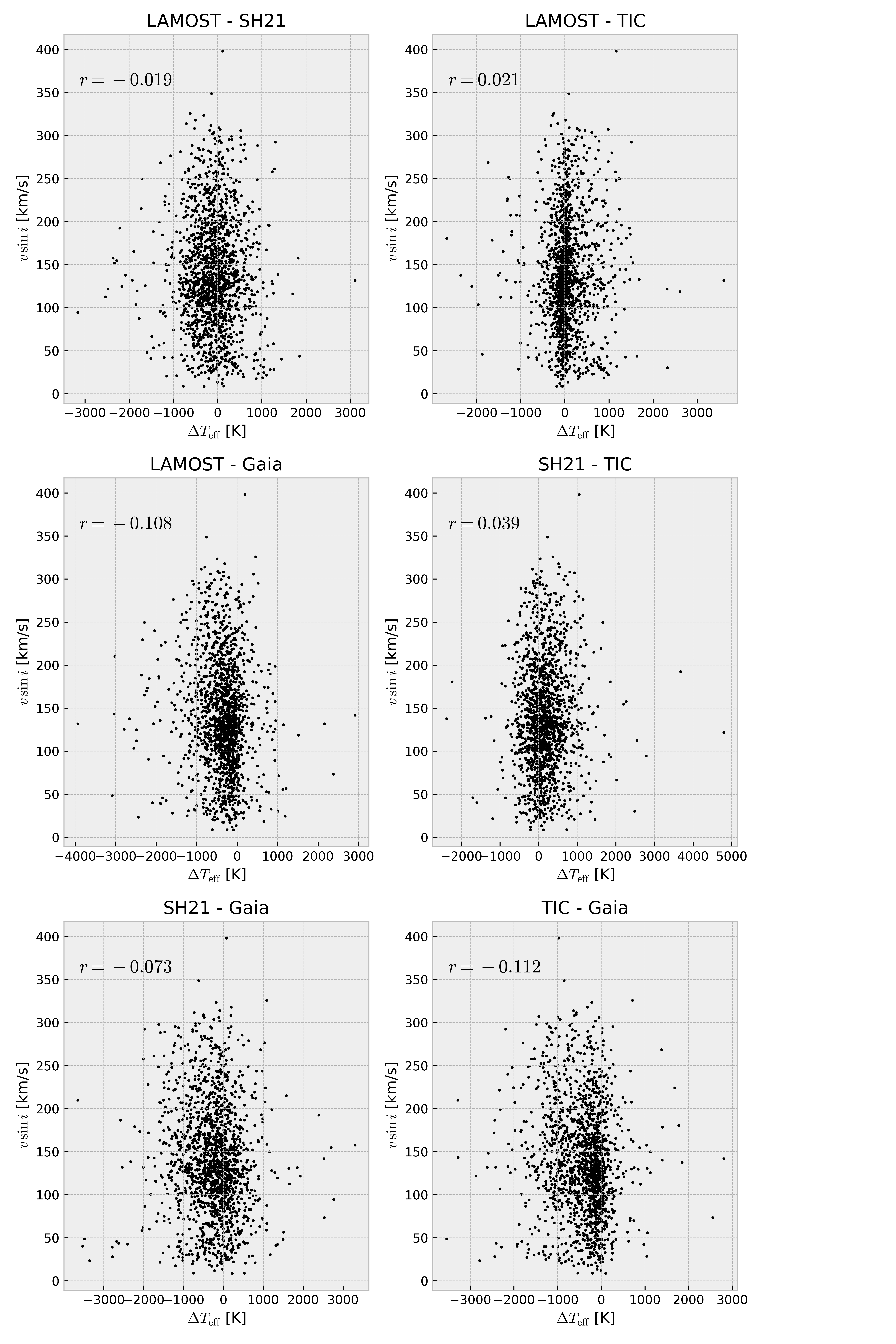}
    \caption{Comparison of the differences in \Teff\, with \vsini\, from the catalogue of \cite{2024ApJS..271....4Z}. We also give the Pearson coefficient (see Section \ref{BC}).}
    \label{fig:teff_fang}
\end{figure}

\begin{figure}[H]
    \includegraphics[width=\textwidth]{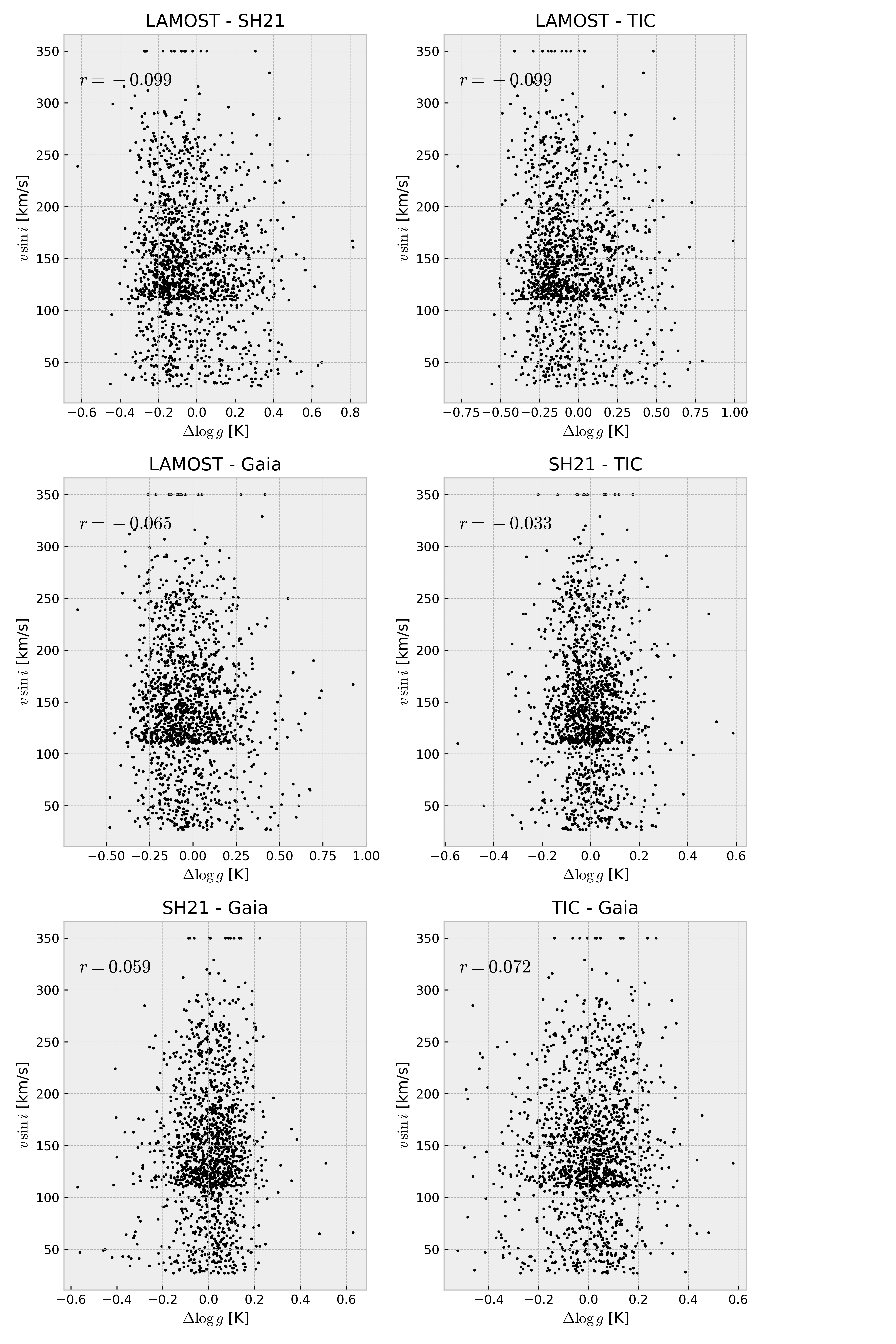}
    \caption{Same as Figure~\ref{fig:teff_fang} but for \logg.}
    \label{fig:logg_fang}
\end{figure}

\begin{figure}[H]
    \includegraphics[width=\textwidth]{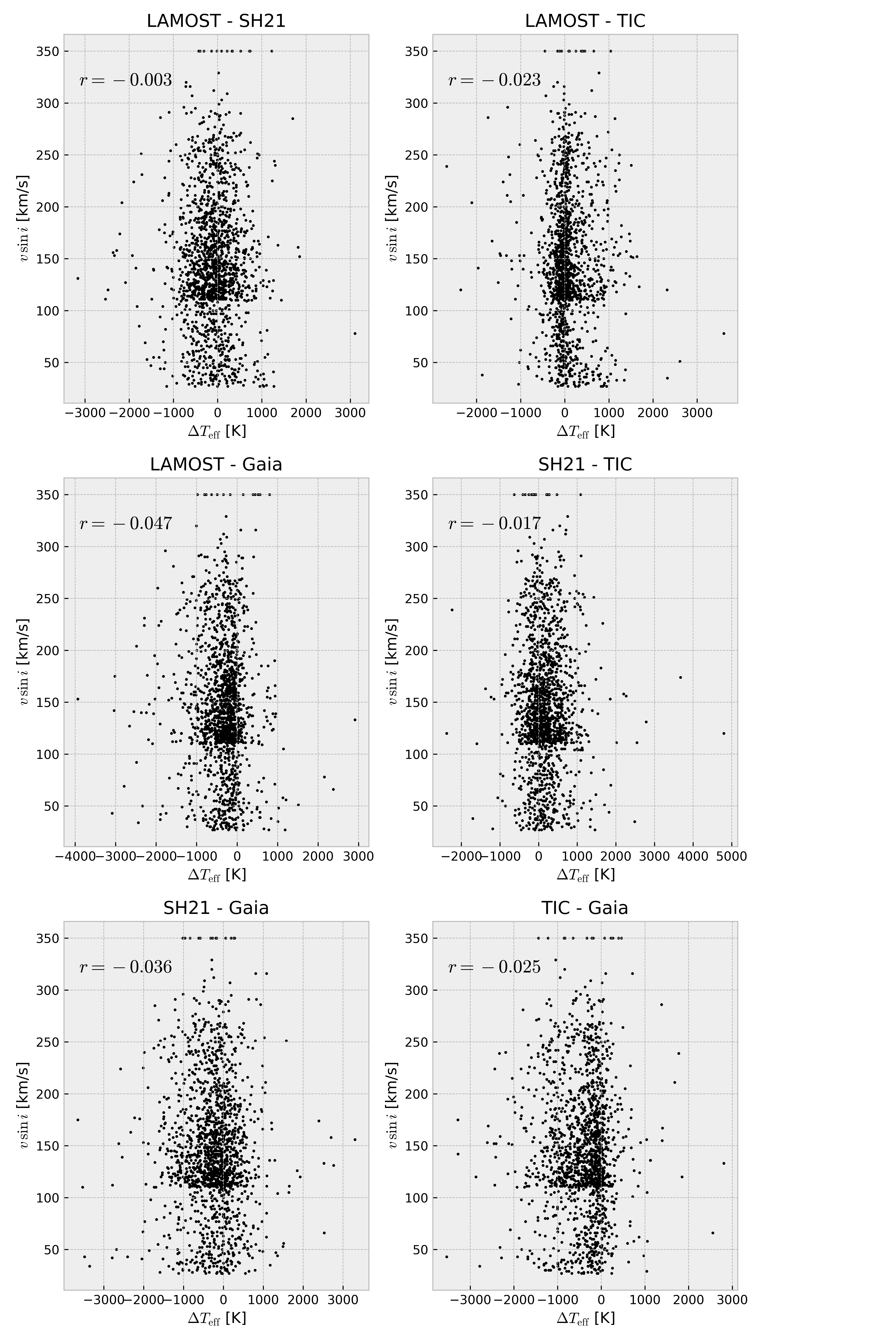}
    \caption{Same as Figure~\ref{fig:teff_fang} but for \vsini\, from \cite{2024A&A...689A.141S}.}
    \label{fig:teff_sun}
\end{figure}

\begin{figure}[H]
   \includegraphics[width=\textwidth]{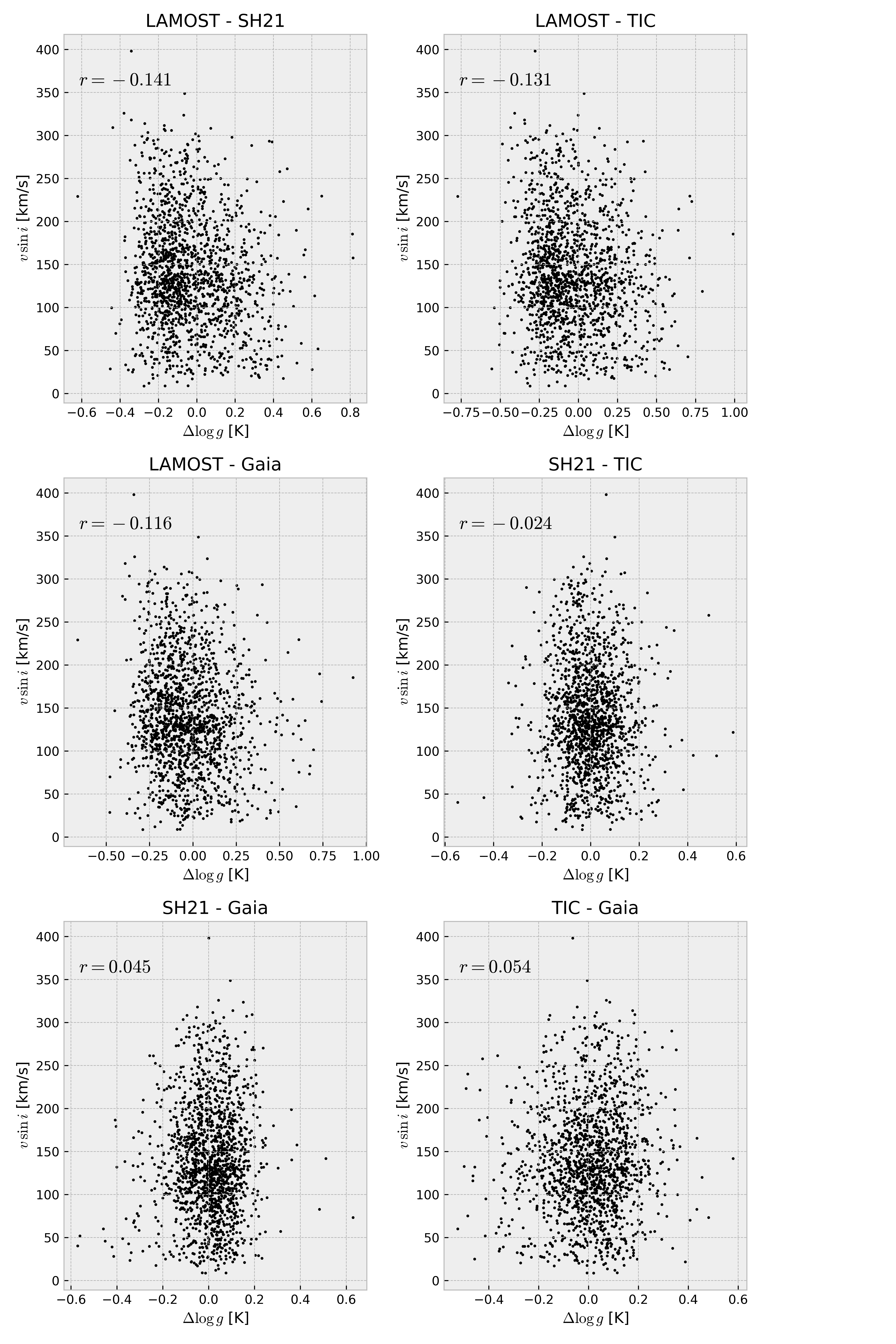}
    \caption{Same as Figure~\ref{fig:teff_sun} but for \logg.}
    \label{fig:logg_sun}
\end{figure}

Additionally, we wanted to see whether there is a dependence on \vsini\, in the differences of \Teff\ and \logg. 
We plotted \vsini\ as a function of differences in the different catalogues. In Figures \ref{fig:teff_fang} and \ref{fig:logg_fang}, one can see the plots for the catalogue of \cite{2024ApJS..271....4Z}, and in Figures \ref{fig:teff_sun} and \ref{fig:logg_sun}, one can see the ones for \cite{2024A&A...689A.141S}. No trend is visible in any of those plots, which is not too surprising, as we assume that \vsini\, would have more of an effect on the star's metallicity than on the other parameters.

\begin{adjustwidth}{-\extralength}{0cm}

\reftitle{References}


\begin{thebibliography}{999}

\end{thebibliography}


\begin{thebibliography}{999}

\bibitem[{Jermyn} et~al.(2022){Jermyn}, {Anders}, {Lecoanet}, and
  {Cantiello}]{2022ApJS..262...19J}
{Jermyn}, A.S.; {Anders}, E.H.; {Lecoanet}, D.; {Cantiello}, M.
\newblock {An Atlas of Convection in Main-sequence Stars}.
\newblock {\em Astrophys. J. Suppl. Ser.} {\bf 2022}, {\em 262},~19.
\newblock {\url{https://doi.org/10.3847/1538-4365/ac7cee}}.

\bibitem[{Pamyatnykh}(1999)]{1999AcA....49..119P}
{Pamyatnykh}, A.A.
\newblock {Pulsational Instability Domains in the Upper Main Sequence}.
\newblock {\em {ActaA}} 
 {\bf 1999}, {\em 49},~119--148.

\bibitem[{D'Antona} et~al.(2002){D'Antona}, {Montalb{\'a}n}, {Kupka}, and
  {Heiter}]{2002ApJ...564L..93D}
{D'Antona}, F.; {Montalb{\'a}n}, J.; {Kupka}, F.; {Heiter}, U.
\newblock {The B{\"o}hm-Vitense Gap: The Role of Turbulent Convection}.
\newblock {\emph{Astrophys. J. Suppl. Ser.}} {\bf 2002}, {\em 564},~L93--L96.
\newblock {\url{https://doi.org/10.1086/338911}}.

\bibitem[{Charbonneau} and {Michaud}(1991)]{1991ApJ...370..693C}
{Charbonneau}, P.; {Michaud}, G.
\newblock {Meridional Circulation and Diffusion in A and Early F Stars}.
\newblock {\em Astrophys. J. } {\bf 1991}, {\em 370},~693.
\newblock {\url{https://doi.org/10.1086/169853}}.

\bibitem[{Moss}(2001)]{2001ASPC..248..305M}
{Moss}, D.
\newblock {Magnetic Fields in the Ap and Bp Stars: A Theoretical Overview}.
\newblock In \emph{Proceedings of the Magnetic Fields Across the Hertzsprung-Russell
  Diagram}; { Astronomical Society of the Pacific Conference Series};{Mathys}, G., {Solanki}, S.K., {Wickramasinghe}, D.T., {San Francisco: Astronomical Society of the Pacific;} 
  2001;
  Volume 248, p.
  305.

\bibitem[{Antoci} et~al.(2019){Antoci}, {Cunha}, {Bowman}, {Murphy}, {Kurtz},
  {Bedding}, {Borre}, {Christophe}, {Daszy{\'n}ska-Daszkiewicz}, {Fox-Machado},
  {Garc{\'\i}a Hern{\'a}ndez}, {Ghasemi}, {Handberg}, {Hansen}, {Hasanzadeh},
  {Houdek}, {Johnston}, {Justesen}, {Kahraman Alicavus}, {Kotysz}, {Latham},
  {Matthews}, {M{\o}nster}, {Niemczura}, {Paunzen}, {S{\'a}nchez Arias},
  {Pigulski}, {Pepper}, {Richey-Yowell}, {Safari}, {Seager}, {Smalley},
  {Shutt}, {S{\'o}dor}, {Su{\'a}rez}, {Tkachenko}, {Wu}, {Zwintz}, {Barcel{\'o}
  Forteza}, {Brunsden}, {Bogn{\'a}r}, {Buzasi}, {Chowdhury}, {De Cat}, {Evans},
  {Guo}, {Guzik}, {Jevtic}, {Lampens}, {Lares Martiz}, {Lovekin}, {Li},
  {Mirouh}, {Mkrtichian}, {Monteiro}, {Nemec}, {Ouazzani}, {Pascual-Granado},
  {Reese}, {Rieutord}, {Rodon}, {Skarka}, {Sowicka}, {Stateva}, {Szab{\'o}},
  and {Weiss}]{2019MNRAS.490.4040A}
{Antoci}, V.; {Cunha}, M.S.; {Bowman}, D.M.; {Murphy}, S.J.; {Kurtz}, D.W.;
  {Bedding}, T.R.; {Borre}, C.C.; {Christophe}, S.;
  {Daszy{\'n}ska-Daszkiewicz}, J.; {Fox-Machado}, L.;  et~al.
\newblock {The first view of {\ensuremath{\delta}} Scuti and
  {\ensuremath{\gamma}} Doradus stars with the TESS mission}.
\newblock {\em Mon. Not. R. Astron. Soc.} {\bf 2019}, {\em 490},~4040--4059.
\newblock {\url{https://doi.org/10.1093/mnras/stz2787}}.

\bibitem[{Walczak} et~al.(2015){Walczak}, {Fontes}, {Colgan}, {Kilcrease}, and
  {Guzik}]{2015A&A...580L...9W}
{Walczak}, P.; {Fontes}, C.J.; {Colgan}, J.; {Kilcrease}, D.P.; {Guzik}, J.A.
\newblock {Wider pulsation instability regions for {\ensuremath{\beta}} Cephei
  and SPB stars calculated using new Los Alamos opacities}.
\newblock {\em Astron. Astrophys.} {\bf 2015}, {\em 580},~L9.
\newblock {\url{https://doi.org/10.1051/0004-6361/201526824}}.

\bibitem[{Maury} and {Pickering}(1897)]{1897AnHar..28....1M}
{Maury}, A.C.; {Pickering}, E.C.
\newblock {Spectra of bright stars photographed with the 11-inch Draper
  Telescope as part of the Henry Draper Memorial.}
\newblock {\em Ann. Harv. Coll. Obs.} {\bf 1897}, {\em
  28},~1--128.

\bibitem[{Alecian}(2023)]{2023MNRAS.519.5913A}
{Alecian}, G.
\newblock {Diffusion velocity of metals in Ap star atmospheres: The effect of
  ambipolar diffusion of H}.
\newblock {\em Mon. Not. R. Astron. Soc.} {\bf 2023}, {\em 519},~5913--5921.
\newblock {\url{https://doi.org/10.1093/mnras/stad034}}.

\bibitem[{ud-Doula} et~al.(2025){ud-Doula}, {Krti{\v{c}}ka}, and
  {Kub{\'a}tov{\'a}}]{2025A&A...694A.270U}
{ud-Doula}, A.; {Krti{\v{c}}ka}, J.; {Kub{\'a}tov{\'a}}, B.
\newblock {Horizontal flows in the atmospheres of chemically peculiar stars}.
\newblock {\em Astron. Astrophys.} {\bf 2025}, {\em 694},~A270.
\newblock {\url{https://doi.org/10.1051/0004-6361/202453189}}.

\bibitem[{Pinsonneault} et~al.(1990){Pinsonneault}, {Kawaler}, and
  {Demarque}]{1990ApJS...74..501P}
{Pinsonneault}, M.H.; {Kawaler}, S.D.; {Demarque}, P.
\newblock {Rotation of Low-Mass Stars: A New Probe of Stellar Evolution}.
\newblock {\em Astrophys. J. Suppl. Ser.} {\bf 1990}, {\em 74},~501.
\newblock {\url{https://doi.org/10.1086/191507}}.

\bibitem[{Abt} et~al.(2002){Abt}, {Levato}, and {Grosso}]{2002ApJ...573..359A}
{Abt}, H.A.; {Levato}, H.; {Grosso}, M.
\newblock {Rotational Velocities of B Stars}.
\newblock {\em Astrophys. J. } {\bf 2002}, {\em 573},~359--365.
\newblock {\url{https://doi.org/10.1086/340590}}.

\bibitem[{Holmberg} et~al.(2007){Holmberg}, {Nordstr{\"o}m}, and
  {Andersen}]{2007A&A...475..519H}
{Holmberg}, J.; {Nordstr{\"o}m}, B.; {Andersen}, J.
\newblock {The Geneva-Copenhagen survey of the Solar neighbourhood II. New uvby
  calibrations and rediscussion of stellar ages, the G dwarf problem,
  age-metallicity diagram, and heating mechanisms of the disk}.
\newblock {\em Astron. Astrophys.} {\bf 2007}, {\em 475},~519--537.
\newblock {\url{https://doi.org/10.1051/0004-6361:20077221}}.

\bibitem[{Teixeira} et~al.(2016){Teixeira}, {Sousa}, {Tsantaki}, {Monteiro},
  {Santos}, and {Israelian}]{2016A&A...595A..15T}
{Teixeira}, G.D.C.; {Sousa}, S.G.; {Tsantaki}, M.; {Monteiro}, M.J.P.F.G.;
  {Santos}, N.C.; {Israelian}, G.
\newblock {New T$_{eff}$ and [Fe/H] spectroscopic calibration for FGK dwarfs
  and GK giants}.
\newblock {\em Astron. Astrophys.} {\bf 2016}, {\em 595},~A15.
\newblock {\url{https://doi.org/10.1051/0004-6361/201525783}}.

\bibitem[{Paunzen} et~al.(2005){Paunzen}, {Schnell}, and
  {Maitzen}]{2005A&A...444..941P}
{Paunzen}, E.; {Schnell}, A.; {Maitzen}, H.M.
\newblock {An empirical temperature calibration for the {\ensuremath{\Delta}} a
  photometric system . I. The B-type stars}.
\newblock {\em Astron. Astrophys.} {\bf 2005}, {\em 444},~941--946.
\newblock {\url{https://doi.org/10.1051/0004-6361:20053546}}.

\bibitem[{Paunzen} et~al.(2006){Paunzen}, {Schnell}, and
  {Maitzen}]{2006A&A...458..293P}
{Paunzen}, E.; {Schnell}, A.; {Maitzen}, H.M.
\newblock {An empirical temperature calibration for the {\ensuremath{\Delta}} a
  photometric system. II. The A-type and mid F-type stars}.
\newblock {\em Astron. Astrophys.} {\bf 2006}, {\em 458},~293--296.
\newblock {\url{https://doi.org/10.1051/0004-6361:20064889}}.

\bibitem[{Cardelli} et~al.(1989){Cardelli}, {Clayton}, and
  {Mathis}]{1989ApJ...345..245C}
{Cardelli}, J.A.; {Clayton}, G.C.; {Mathis}, J.S.
\newblock {The Relationship between Infrared, Optical, and Ultraviolet
  Extinction}.
\newblock {\em Astrophys. J. } {\bf 1989}, {\em 345},~245.
\newblock {\url{https://doi.org/10.1086/167900}}.

\bibitem[{McDonald} et~al.(2024){McDonald}, {Zijlstra}, {Cox}, {Alexander},
  {Csukai}, {Ramkumar}, and {Hollings}]{2024RASTI...3...89M}
{McDonald}, I.; {Zijlstra}, A.A.; {Cox}, N.L.J.; {Alexander}, E.L.; {Csukai},
  A.; {Ramkumar}, R.; {Hollings}, A.
\newblock {PYSSED: An automated method of collating and fitting stellar
  spectral energy distributions}.
\newblock {\em RAS Tech. Instruments} {\bf 2024}, {\em 3},~89--107.
\newblock {\url{https://doi.org/10.1093/rasti/rzae005}}.

\bibitem[{Marigo} et~al.(2001){Marigo}, {Girardi}, {Chiosi}, and
  {Wood}]{2001A&A...371..152M}
{Marigo}, P.; {Girardi}, L.; {Chiosi}, C.; {Wood}, P.R.
\newblock {Zero-metallicity stars. I. Evolution at constant mass}.
\newblock {\em Astron. Astrophys.} {\bf 2001}, {\em 371},~152--173.
\newblock {\url{https://doi.org/10.1051/0004-6361:20010309}}.

\bibitem[{Allende Prieto} et~al.(2004){Allende Prieto}, {Barklem}, {Lambert},
  and {Cunha}]{2004A&A...420..183A}
{Allende Prieto}, C.; {Barklem}, P.S.; {Lambert}, D.L.; {Cunha}, K.
\newblock {S$^{4}$N: A spectroscopic survey of stars in the solar neighborhood.
  The Nearest 15 pc}.
\newblock {\em Astron. Astrophys.} {\bf 2004}, {\em 420},~183--205.
\newblock {\url{https://doi.org/10.1051/0004-6361:20035801}}.

\bibitem[{Gebran} et~al.(2010){Gebran}, {Vick}, {Monier}, and
  {Fossati}]{2010A&A...523A..71G}
{Gebran}, M.; {Vick}, M.; {Monier}, R.; {Fossati}, L.
\newblock {Chemical composition of A and F dwarfs members of the Hyades open
  cluster}.
\newblock {\em Astron. Astrophys.} {\bf 2010}, {\em 523},~A71.
\newblock {\url{https://doi.org/10.1051/0004-6361/200913273}}.

\bibitem[{Shejeelammal} et~al.(2024){Shejeelammal}, {Mel{\'e}ndez}, {Rathsam},
  and {Martos}]{2024A&A...690A.107S}
{Shejeelammal}, J.; {Mel{\'e}ndez}, J.; {Rathsam}, A.; {Martos}, G.
\newblock {The [Y/Mg] chemical clock in the Galactic disk: The influence of
  metallicity and the Galactic population in the solar neighbourhood}.
\newblock {\em Astron. Astrophys.} {\bf 2024}, {\em 690},~A107.
\newblock {\url{https://doi.org/10.1051/0004-6361/202449669}}.

\bibitem[{Netopil} et~al.(2022){Netopil}, {Oralhan}, {{\c{C}}akmak}, {Michel},
  and {Karata{\c{s}}}]{2022MNRAS.509..421N}
{Netopil}, M.; {Oralhan}, {\.I}.A.; {{\c{C}}akmak}, H.; {Michel}, R.;
  {Karata{\c{s}}}, Y.
\newblock {The Galactic metallicity gradient shown by open clusters in the
  light of radial migration}.
\newblock {\em Mon. Not. R. Astron. Soc.} {\bf 2022}, {\em 509},~421--439.
\newblock {\url{https://doi.org/10.1093/mnras/stab2961}}.

\bibitem[{Zhao} et~al.(2012){Zhao}, {Zhao}, {Chu}, {Jing}, and
  {Deng}]{2012RAA....12..723Z}
{Zhao}, G.; {Zhao}, Y.H.; {Chu}, Y.Q.; {Jing}, Y.P.; {Deng}, L.C.
\newblock {LAMOST spectral survey {\textemdash} An overview}.
\newblock {\em Res. Astron. Astrophys.} {\bf 2012}, {\em
  12},~723--734.
\newblock {\url{https://doi.org/10.1088/1674-4527/12/7/002}}.

\bibitem[{Gaia Collaboration} et~al.(2023){Gaia Collaboration}, {Creevey},
  {Sarro}, {Lobel}, {Pancino}, {Andrae}, {Smart}, {Clementini}, {Heiter},
  {Korn}, {Fouesneau}, {Fr{\'e}mat}, {De Angeli}, {Vallenari}, {Harrison},
  {Th{\'e}venin}, {Reyl{\'e}}, {Sordo}, {Garofalo}, {Brown}, {Eyer}, {Prusti},
  {de Bruijne}, {Arenou}, {Babusiaux}, {Biermann}, {Ducourant}, {Evans},
  {Guerra}, {Hutton}, {Jordi}, {Klioner}, {Lammers}, {Lindegren}, {Luri},
  {Mignard}, {Panem}, {Pourbaix}, {Randich}, {Sartoretti}, {Soubiran}, {Tanga},
  {Walton}, {Bailer-Jones}, {Bastian}, {Drimmel}, {Jansen}, {Katz}, {Lattanzi},
  {van Leeuwen}, {Bakker}, {Cacciari}, {Casta{\~n}eda}, {Fabricius},
  {Galluccio}, {Guerrier}, {Masana}, {Messineo}, {Mowlavi}, {Nicolas},
  {Nienartowicz}, {Pailler}, {Panuzzo}, {Riclet}, {Roux}, {Seabroke},
  {Gracia-Abril}, {Portell}, {Teyssier}, {Altmann}, {Audard}, {Bellas-Velidis},
  {Benson}, {Berthier}, {Blomme}, {Burgess}, {Busonero}, {Busso},
  {C{\'a}novas}, {Carry}, {Cellino}, {Cheek}, {Damerdji}, {Davidson}, {de
  Teodoro}, {Nu{\~n}ez Campos}, {Delchambre}, {Dell'Oro}, {Esquej},
  {Fern{\'a}ndez-Hern{\'a}ndez}, {Fraile}, {Garabato}, {Garc{\'\i}a-Lario},
  {Gosset}, {Haigron}, {Halbwachs}, {Hambly}, {Hern{\'a}ndez}, {Hestroffer},
  {Hodgkin}, {Holl}, {Jan{\ss}en}, {Jevardat de Fombelle}, {Jordan},
  {Krone-Martins}, {Lanzafame}, {L{\"o}ffler}, {Marchal}, {Marrese},
  {Moitinho}, {Muinonen}, {Osborne}, {Pauwels}, {Recio-Blanco}, {Riello},
  {Rimoldini}, {Roegiers}, {Rybizki}, {Siopis}, {Smith}, {Sozzetti}, {Utrilla},
  {van Leeuwen}, {Abbas}, {{\'A}brah{\'a}m}, {Abreu Aramburu}, {Aerts},
  {Aguado}, {Ajaj}, {Aldea-Montero}, {Altavilla}, {{\'A}lvarez}, {Alves},
  {Anders}, {Anderson}, {Anglada Varela}, {Antoja}, {Baines}, {Baker},
  {Balaguer-N{\'u}{\~n}ez}, {Balbinot}, {Balog}, {Barache}, {Barbato},
  {Barros}, {Barstow}, {Bartolom{\'e}}, {Bassilana}, {Bauchet}, {Becciani},
  {Bellazzini}, {Berihuete}, {Bernet}, {Bertone}, {Bianchi}, {Binnenfeld},
  {Blanco-Cuaresma}, {Boch}, {Bombrun}, {Bossini}, {Bouquillon}, {Bragaglia},
  {Bramante}, {Breedt}, {Bressan}, {Brouillet}, {Brugaletta}, {Bucciarelli},
  {Burlacu}, {Butkevich}, {Buzzi}, {Caffau}, {Cancelliere}, {Cantat-Gaudin},
  {Carballo}, {Carlucci}, {Carnerero}, {Carrasco}, {Casamiquela}, {Castellani},
  {Castro-Ginard}, {Chaoul}, {Charlot}, {Chemin}, {Chiaramida}, {Chiavassa},
  {Chornay}, {Comoretto}, {Contursi}, {Cooper}, {Cornez}, {Cowell}, {Crifo},
  {Cropper}, {Crosta}, {Crowley}, {Dafonte}, {Dapergolas}, {David}, {de
  Laverny}, {De Luise}, {De March}, {De Ridder}, {de Souza}, {de Torres}, {del
  Peloso}, {del Pozo}, {Delbo}, {Delgado}, {Delisle}, {Demouchy},
  {Dharmawardena}, {Di Matteo}, {Diakite}, {Diener}, {Distefano}, {Dolding},
  {Enke}, {Fabre}, {Fabrizio}, {Faigler}, {Fedorets}, {Fernique}, {Figueras},
  {Fournier}, {Fouron}, {Fragkoudi}, {Gai}, {Garcia-Gutierrez},
  {Garcia-Reinaldos}, {Garc{\'\i}a-Torres}, {Gavel}, {Gavras}, {Gerlach},
  {Geyer}, {Giacobbe}, {Gilmore}, {Girona}, {Giuffrida}, {Gomel}, {Gomez},
  {Gonz{\'a}lez-N{\'u}{\~n}ez}, {Gonz{\'a}lez-Santamar{\'\i}a},
  {Gonz{\'a}lez-Vidal}, {Granvik}, {Guillout}, {Guiraud},
  {Guti{\'e}rrez-S{\'a}nchez}, {Guy}, {Hatzidimitriou}, {Hauser}, {Haywood},
  {Helmer}, {Helmi}, {Hilger}, {Sarmiento}, {Hidalgo}, {H{\l}adczuk}, {Hobbs},
  {Holland}, {Huckle}, {Jardine}, {Jasniewicz}, {Jean-Antoine Piccolo},
  {Jim{\'e}nez-Arranz}, {Juaristi Campillo}, {Julbe}, {Karbevska}, {Kervella},
  {Khanna}, {Kordopatis}, {K{\'o}sp{\'a}l}, {Kostrzewa-Rutkowska},
  {Kruszy{\'n}ska}, {Kun}, {Laizeau}, {Lambert}, {Lanza}, {Lasne}, {Le
  Campion}, {Lebreton}, {Lebzelter}, {Leccia}, {Leclerc}, {Lecoeur-Taibi},
  {Liao}, {Licata}, {Lindstr{\o}m}, {Lister}, {Livanou}, {Lorca}, {Loup},
  {Madrero Pardo}, {Magdaleno Romeo}, {Managau}, {Mann}, {Manteiga},
  {Marchant}, {Marconi}, {Marcos}, {Marcos Santos}, {Mar{\'\i}n Pina},
  {Marinoni}, {Marocco}, {Marshall}, {Martin Polo}, {Mart{\'\i}n-Fleitas},
  {Marton}, {Mary}, {Masip}, {Massari}, {Mastrobuono-Battisti}, {Mazeh},
  {McMillan}, {Messina}, {Michalik}, {Millar}, {Mints}, {Molina}, {Molinaro},
  {Moln{\'a}r}, {Monari}, {Mongui{\'o}}, {Montegriffo}, {Montero}, {Mor},
  {Mora}, {Morbidelli}, {Morel}, {Morris}, {Muraveva}, {Murphy}, {Musella},
  {Nagy}, {Noval}, {Oca{\~n}a}, {Ogden}, {Ordenovic}, {Osinde}, {Pagani},
  {Pagano}, {Palaversa}, {Palicio}, {Pallas-Quintela}, {Panahi},
  {Payne-Wardenaar}, {Pe{\~n}alosa Esteller}, {Penttil{\"a}}, {Pichon},
  {Piersimoni}, {Pineau}, {Plachy}, {Plum}, {Poggio}, {Pr{\v{s}}a}, {Pulone},
  {Racero}, {Ragaini}, {Rainer}, {Raiteri}, {Ramos}, {Ramos-Lerate}, {Re
  Fiorentin}, {Regibo}, {Richards}, {Rios Diaz}, {Ripepi}, {Riva}, {Rix},
  {Rixon}, {Robichon}, {Robin}, {Robin}, {Roelens}, {Rogues}, {Rohrbasser},
  {Romero-G{\'o}mez}, {Rowell}, {Royer}, {Ruz Mieres}, {Rybicki}, {Sadowski},
  {S{\'a}ez N{\'u}{\~n}ez}, {Sagrist{\`a} Sell{\'e}s}, {Sahlmann}, {Salguero},
  {Samaras}, {Sanchez Gimenez}, {Sanna}, {Santove{\~n}a}, {Sarasso},
  {Schultheis}, {Sciacca}, {Segol}, {Segovia}, {S{\'e}gransan}, {Semeux},
  {Shahaf}, {Siddiqui}, {Siebert}, {Siltala}, {Silvelo}, {Slezak}, {Slezak},
  {Snaith}, {Solano}, {Solitro}, {Souami}, {Souchay}, {Spagna}, {Spina},
  {Spoto}, {Steele}, {Steidelm{\"u}ller}, {Stephenson}, {S{\"u}veges},
  {Surdej}, {Szabados}, {Szegedi-Elek}, {Taris}, {Taylor}, {Teixeira},
  {Tolomei}, {Tonello}, {Torra}, {Torra}, {Torralba Elipe}, {Trabucchi},
  {Tsounis}, {Turon}, {Ulla}, {Unger}, {Vaillant}, {van Dillen}, {van Reeven},
  {Vanel}, {Vecchiato}, {Viala}, {Vicente}, {Voutsinas}, {Weiler}, {Wevers},
  {Wyrzykowski}, {Yoldas}, {Yvard}, {Zhao}, {Zorec}, {Zucker}, and
  {Zwitter}]{2023A&A...674A..39G}
{Gaia Collaboration}; {Creevey}, O.L.; {Sarro}, L.M.; {Lobel}, A.; {Pancino},
  E.; {Andrae}, R.; {Smart}, R.L.; {Clementini}, G.; {Heiter}, U.; {Korn},
  A.J.;  et~al.
\newblock {Gaia Data Release 3. A golden sample of astrophysical parameters}.
\newblock {\em Astron. Astrophys.} {\bf 2023}, {\em 674},~A39.
\newblock {\url{https://doi.org/10.1051/0004-6361/202243800}}.

\bibitem[{Gray} and {Corbally}(2014)]{2014AJ....147...80G}
{Gray}, R.O.; {Corbally}, C.J.
\newblock {An Expert Computer Program for Classifying Stars on the MK Spectral
  Classification System}.
\newblock {\em Astron. J.} {\bf 2014}, {\em 147},~80.
\newblock {\url{https://doi.org/10.1088/0004-6256/147/4/80}}.

\bibitem[{H{\"u}mmerich} et~al.(2020){H{\"u}mmerich}, {Paunzen}, and
  {Bernhard}]{2020A&A...640A..40H}
{H{\"u}mmerich}, S.; {Paunzen}, E.; {Bernhard}, K.
\newblock {A plethora of new, magnetic chemically peculiar stars from LAMOST
  DR4}.
\newblock {\em Astron. Astrophys.} {\bf 2020}, {\em 640},~A40.
\newblock {\url{https://doi.org/10.1051/0004-6361/202037750}}.

\bibitem[{Anders} et~al.(2019){Anders}, {Khalatyan}, {Chiappini}, {Queiroz},
  {Santiago}, {Jordi}, {Girardi}, {Brown}, {Matijevi{\v{c}}}, {Monari},
  {Cantat-Gaudin}, {Weiler}, {Khan}, {Miglio}, {Carrillo}, {Romero-G{\'o}mez},
  {Minchev}, {de Jong}, {Antoja}, {Ramos}, {Steinmetz}, and
  {Enke}]{2019AandA...628A..94A}
{Anders}, F.; {Khalatyan}, A.; {Chiappini}, C.; {Queiroz}, A.B.; {Santiago},
  B.X.; {Jordi}, C.; {Girardi}, L.; {Brown}, A.G.A.; {Matijevi{\v{c}}}, G.;
  {Monari}, G.;  et~al.
\newblock {Photo-astrometric distances, extinctions, and astrophysical
  parameters for Gaia DR2 stars brighter than G = 18}.
\newblock {\em Astron. Astrophys.} {\bf 2019}, {\em 628},~A94.
\newblock {\url{https://doi.org/10.1051/0004-6361/201935765}}.

\bibitem[{Anders} et~al.(2022){Anders}, {Khalatyan}, {Queiroz}, {Chiappini},
  {Ard{\`e}vol}, {Casamiquela}, {Figueras}, {Jim{\'e}nez-Arranz}, {Jordi},
  {Mongui{\'o}}, {Romero-G{\'o}mez}, {Altamirano}, {Antoja}, {Assaad},
  {Cantat-Gaudin}, {Castro-Ginard}, {Enke}, {Girardi}, {Guiglion}, {Khan},
  {Luri}, {Miglio}, {Minchev}, {Ramos}, {Santiago}, and
  {Steinmetz}]{2022AandA...658A..91A}
{Anders}, F.; {Khalatyan}, A.; {Queiroz}, A.B.A.; {Chiappini}, C.;
  {Ard{\`e}vol}, J.; {Casamiquela}, L.; {Figueras}, F.; {Jim{\'e}nez-Arranz},
  {\'O}.; {Jordi}, C.; {Mongui{\'o}}, M.;  et~al.
\newblock {Photo-astrometric distances, extinctions, and astrophysical
  parameters for Gaia EDR3 stars brighter than G = 18.5}.
\newblock {\em Astron. Astrophys.} {\bf 2022}, {\em 658},~A91.
\newblock {\url{https://doi.org/10.1051/0004-6361/202142369}}.

\bibitem[{Magnier} et~al.(2013){Magnier}, {Schlafly}, {Finkbeiner}, {Juric},
  {Tonry}, {Burgett}, {Chambers}, {Flewelling}, {Kaiser}, {Kudritzki},
  {Morgan}, {Price}, {Sweeney}, and {Stubbs}]{2013ApJS..205...20M}
{Magnier}, E.A.; {Schlafly}, E.; {Finkbeiner}, D.; {Juric}, M.; {Tonry}, J.L.;
  {Burgett}, W.S.; {Chambers}, K.C.; {Flewelling}, H.A.; {Kaiser}, N.;
  {Kudritzki}, R.P.;  et~al.
\newblock {The Pan-STARRS 1 Photometric Reference Ladder, Release 12.01}.
\newblock {\em Astrophys. J. Suppl. Ser.} {\bf 2013}, {\em 205},~20.
\newblock {\url{https://doi.org/10.1088/0067-0049/205/2/20}}.

\bibitem[{Skrutskie} et~al.(2006){Skrutskie}, {Cutri}, {Stiening}, {Weinberg},
  {Schneider}, {Carpenter}, {Beichman}, {Capps}, {Chester}, {Elias}, {Huchra},
  {Liebert}, {Lonsdale}, {Monet}, {Price}, {Seitzer}, {Jarrett}, {Kirkpatrick},
  {Gizis}, {Howard}, {Evans}, {Fowler}, {Fullmer}, {Hurt}, {Light}, {Kopan},
  {Marsh}, {McCallon}, {Tam}, {Van Dyk}, and {Wheelock}]{2006AJ....131.1163S}
{Skrutskie}, M.F.; {Cutri}, R.M.; {Stiening}, R.; {Weinberg}, M.D.;
  {Schneider}, S.; {Carpenter}, J.M.; {Beichman}, C.; {Capps}, R.; {Chester},
  T.; {Elias}, J.;  et~al.
\newblock {The Two Micron All Sky Survey (2MASS)}.
\newblock {\em Astron. J.} {\bf 2006}, {\em 131},~1163--1183.
\newblock {\url{https://doi.org/10.1086/498708}}.

\bibitem[{Cutri} and {et al.}(2013)]{2013yCat.2328....0C}
{Cutri}, R.M.; Wright, E.L.; Conrow, T.; Fowler, J.W.; Eisenhardt, P.R.M.; Grillmair, C.; {Kirkpatrick}, J.D.;  {Masci}, F.;  {McCallon}, H.L.;  {Wheelock}, S.L.; {et al.}
\newblock {\emph{VizieR Online Data Catalog: AllWISE Data Release (Cutri+ 2013)}};
\newblock { {VizieR Online Data Catalog:} 
 } { 2013}; p. II/328.

\bibitem[{Onken} et~al.(2019){Onken}, {Wolf}, {Bessell}, {Chang}, {Da Costa},
  {Luvaul}, {Mackey}, {Schmidt}, and {Shao}]{2019PASA...36...33O}
{Onken}, C.A.; {Wolf}, C.; {Bessell}, M.S.; {Chang}, S.W.; {Da Costa}, G.S.;
  {Luvaul}, L.C.; {Mackey}, D.; {Schmidt}, B.P.; {Shao}, L.
\newblock {SkyMapper Southern Survey: Second data release (DR2)}.
\newblock {\em Publ. Astron. Soc. Aust.} {\bf 2019}, {\em 36},~e033.
\newblock {\url{https://doi.org/10.1017/pasa.2019.27}}.

\bibitem[{Queiroz} et~al.(2018){Queiroz}, {Anders}, {Santiago}, {Chiappini},
  {Steinmetz}, {Dal Ponte}, {Stassun}, {da Costa}, {Maia}, {Crestani}, {Beers},
  {Fern{\'a}ndez-Trincado}, {Garc{\'\i}a-Hern{\'a}ndez}, {Roman-Lopes}, and
  {Zamora}]{2018MNRAS.476.2556Q}
{Queiroz}, A.B.A.; {Anders}, F.; {Santiago}, B.X.; {Chiappini}, C.;
  {Steinmetz}, M.; {Dal Ponte}, M.; {Stassun}, K.G.; {da Costa}, L.N.; {Maia},
  M.A.G.; {Crestani}, J.;  et~al.
\newblock {StarHorse: A Bayesian tool for determining stellar masses, ages,
  distances, and extinctions for field stars}.
\newblock {\em Mon. Not. R. Astron. Soc.} {\bf 2018}, {\em 476},~2556--2583.
\newblock {\url{https://doi.org/10.1093/mnras/sty330}}.

\bibitem[{Fouesneau} et~al.(2023){Fouesneau}, {Fr{\'e}mat}, {Andrae}, {Korn},
  {Soubiran}, {Kordopatis}, {Vallenari}, {Heiter}, {Creevey}, {Sarro}, {de
  Laverny}, {Lanzafame}, {Lobel}, {Sordo}, {Rybizki}, {Slezak}, {{\'A}lvarez},
  {Drimmel}, {Garabato}, {Delchambre}, {Bailer-Jones}, {Hatzidimitriou},
  {Lorca}, {Le Fustec}, {Pailler}, {Mary}, {Robin}, {Utrilla}, {Abreu
  Aramburu}, {Bakker}, {Bellas-Velidis}, {Bijaoui}, {Blomme}, {Bouret},
  {Brouillet}, {Brugaletta}, {Burlacu}, {Carballo}, {Casamiquela}, {Chaoul},
  {Chiavassa}, {Contursi}, {Cooper}, {Dafonte}, {Demouchy}, {Dharmawardena},
  {Garc{\'\i}a-Lario}, {Garc{\'\i}a-Torres}, {Gomez},
  {Gonz{\'a}lez-Santamar{\'\i}a}, {Jean-Antoine Piccolo}, {Kontizas},
  {Lebreton}, {Licata}, {Lindstr{\o}m}, {Livanou}, {Magdaleno Romeo},
  {Manteiga}, {Marocco}, {Martayan}, {Marshall}, {Nicolas}, {Ordenovic},
  {Palicio}, {Pallas-Quintela}, {Pichon}, {Poggio}, {Recio-Blanco}, {Riclet},
  {Santove{\~n}a}, {Schultheis}, {Segol}, {Silvelo}, {Smart}, {S{\"u}veges},
  {Th{\'e}venin}, {Torralba Elipe}, {Ulla}, {van Dillen}, {Zhao}, and
  {Zorec}]{2023AandA...674A..28F}
{Fouesneau}, M.; {Fr{\'e}mat}, Y.; {Andrae}, R.; {Korn}, A.J.; {Soubiran}, C.;
  {Kordopatis}, G.; {Vallenari}, A.; {Heiter}, U.; {Creevey}, O.L.; {Sarro},
  L.M.;  et~al.
\newblock {Gaia Data Release 3. Apsis. II. Stellar parameters}.
\newblock {\em Astron. Astrophys.} {\bf 2023}, {\em 674},~A28.
\newblock {\url{https://doi.org/10.1051/0004-6361/202243919}}.

\bibitem[{Li} et~al.(2022){Li}, {Zeng}, {Wang}, {Du}, {Kong}, and
  {Liao}]{2022MNRAS.514.4588L}
{Li}, X.; {Zeng}, S.; {Wang}, Z.; {Du}, B.; {Kong}, X.; {Liao}, C.
\newblock {Estimating atmospheric parameters from LAMOST low-resolution spectra
  with low SNR}.
\newblock {\em Mon. Not. R. Astron. Soc.} {\bf 2022}, {\em 514},~4588--4600.
\newblock {\url{https://doi.org/10.1093/mnras/stac1625}}.

\bibitem[{Stassun} et~al.(2019){Stassun}, {Oelkers}, {Paegert}, {Torres},
  {Pepper}, {De Lee}, {Collins}, {Latham}, {Muirhead}, {Chittidi},
  {Rojas-Ayala}, {Fleming}, {Rose}, {Tenenbaum}, {Ting}, {Kane}, {Barclay},
  {Bean}, {Brassuer}, {Charbonneau}, {Ge}, {Lissauer}, {Mann}, {McLean},
  {Mullally}, {Narita}, {Plavchan}, {Ricker}, {Sasselov}, {Seager}, {Sharma},
  {Shiao}, {Sozzetti}, {Stello}, {Vanderspek}, {Wallace}, and
  {Winn}]{2019AJ....158..138S}
{Stassun}, K.G.; {Oelkers}, R.J.; {Paegert}, M.; {Torres}, G.; {Pepper}, J.;
  {De Lee}, N.; {Collins}, K.; {Latham}, D.W.; {Muirhead}, P.S.; {Chittidi},
  J.;  et~al.
\newblock {The Revised TESS Input Catalog and Candidate Target List}.
\newblock {\em Astron. J.} {\bf 2019}, {\em 158},~138.
\newblock {\url{https://doi.org/10.3847/1538-3881/ab3467}}.

\bibitem[{Husser} et~al.(2013){Husser}, {Wende-von Berg}, {Dreizler},
  {Homeier}, {Reiners}, {Barman}, and {Hauschildt}]{2013A&A...553A...6H}
{Husser}, T.O.; {Wende-von Berg}, S.; {Dreizler}, S.; {Homeier}, D.; {Reiners},
  A.; {Barman}, T.; {Hauschildt}, P.H.
\newblock {A new extensive library of PHOENIX stellar atmospheres and synthetic
  spectra}.
\newblock {\em Astron. Astrophys.} {\bf 2013}, {\em 553},~A6.
\newblock {\url{https://doi.org/10.1051/0004-6361/201219058}}.

\bibitem[{Stassun} et~al.(2018){Stassun}, {Oelkers}, {Pepper}, {Paegert}, {De
  Lee}, {Torres}, {Latham}, {Charpinet}, {Dressing}, {Huber}, {Kane},
  {L{\'e}pine}, {Mann}, {Muirhead}, {Rojas-Ayala}, {Silvotti}, {Fleming},
  {Levine}, and {Plavchan}]{2018AJ....156..102S}
{Stassun}, K.G.; {Oelkers}, R.J.; {Pepper}, J.; {Paegert}, M.; {De Lee}, N.;
  {Torres}, G.; {Latham}, D.W.; {Charpinet}, S.; {Dressing}, C.D.; {Huber}, D.;
   et~al.
\newblock {The TESS Input Catalog and Candidate Target List}.
\newblock {\em Astron. J.} {\bf 2018}, {\em 156},~102.
\newblock {\url{https://doi.org/10.3847/1538-3881/aad050}}.

\bibitem[{Sun} and {Chiappini}(2024)]{2024A&A...689A.141S}
{Sun}, W.; {Chiappini}, C.
\newblock {Exploring the stellar rotation of early-type stars in the LAMOST
  medium-resolution survey: III. Evolution}.
\newblock {\em Astron. Astrophys.} {\bf 2024}, {\em 689},~A141.
\newblock {\url{https://doi.org/10.1051/0004-6361/202450628}}.

\bibitem[{Zhang} et~al.(2020){Zhang}, {Liu}, and {Deng}]{2020ApJS..246....9Z}
{Zhang}, B.; {Liu}, C.; {Deng}, L.C.
\newblock {Deriving the Stellar Labels of LAMOST Spectra with the Stellar LAbel
  Machine (SLAM)}.
\newblock {\em Astrophys. J. Suppl. Ser.} {\bf 2020}, {\em 246},~9.
\newblock {\url{https://doi.org/10.3847/1538-4365/ab55ef}}.

\bibitem[{Castelli}(2005)]{2005MSAIS...8...25C}
{Castelli}, F.
\newblock {ATLAS12: How to use it}.
\newblock {\em Mem. Della Soc. Astron. Ital. Suppl.} {\bf
  2005}, {\em 8},~25.

\bibitem[{Zuo} et~al.(2024){Zuo}, {Luo}, {Du}, {Li}, {Jones}, {Song}, {Kong},
  and {Guo}]{2024ApJS..271....4Z}
{Zuo}, F.; {Luo}, A.L.; {Du}, B.; {Li}, Y.; {Jones}, H.R.A.; {Song}, Y.h.;
  {Kong}, X.; {Guo}, Y.x.
\newblock {Projected Rotational Velocities for LAMOST Stars with Effective
  Temperatures Lower than 9000 K}.
\newblock {\em Astrophys. J. Suppl. Ser.} {\bf 2024}, {\em 271},~4.
\newblock {\url{https://doi.org/10.3847/1538-4365/ad1eeb}}.

\bibitem[{Majewski}(2015)]{2015mwss.confE..47M}
{Majewski}, S.
\newblock {The APO Galactic Evolution Experiment (APOGEE)}.
\newblock In \emph{Proceedings of the The Milky Way and its Stars: Stellar
  Astrophysics, Galactic Archaeology, and Stellar {Populations}
};  2015; p.~47.

\bibitem[{Bensby} et~al.(2014){Bensby}, {Feltzing}, and
  {Oey}]{2014A&A...562A..71B}
{Bensby}, T.; {Feltzing}, S.; {Oey}, M.S.
\newblock {Exploring the Milky Way stellar disk. A detailed elemental abundance
  study of 714 F and G dwarf stars in the solar neighbourhood}.
\newblock {\em Astron. Astrophys.} {\bf 2014}, {\em 562},~A71.
\newblock {\url{https://doi.org/10.1051/0004-6361/201322631}}.

\bibitem[{Straizys} et~al.(1982){Straizys}, {Jodinskiene}, and
  {Kurliene}]{1982VilOB..60...16S}
{Straizys}, V.; {Jodinskiene}, E.; {Kurliene}, G.
\newblock {The calibration of the Vilnius photometric system in temperatures
  and gravities}.
\newblock {\em Vilnius Astron. Obs. Biul.} {\bf 1982}, {\em
  60},~16.

\bibitem[{Karata{\c{s}}} and {Schuster}(2006)]{2006MNRAS.371.1793K}
{Karata{\c{s}}}, Y.; {Schuster}, W.J.
\newblock {Metallicity and absolute magnitude calibrations for UBV photometry}.
\newblock {\em Mon. Not. R. Astron. Soc.} {\bf 2006}, {\em 371},~1793--1812.
\newblock {\url{https://doi.org/10.1111/j.1365-2966.2006.10800.x}}.

\bibitem[{Paunzen} and {Netopil}(2006)]{2006MNRAS.371.1641P}
{Paunzen}, E.; {Netopil}, M.
\newblock {On the current status of open-cluster parameters}.
\newblock {\em Mon. Not. R. Astron. Soc.} {\bf 2006}, {\em 371},~1641--1647.
\newblock {\url{https://doi.org/10.1111/j.1365-2966.2006.10783.x}}.

\bibitem[{Holmberg} et~al.(2009){Holmberg}, {Nordstr{\"o}m}, and
  {Andersen}]{2009A&A...501..941H}
{Holmberg}, J.; {Nordstr{\"o}m}, B.; {Andersen}, J.
\newblock {The Geneva-Copenhagen survey of the solar neighbourhood. III.
  Improved distances, ages, and kinematics}.
\newblock {\em Astron. Astrophys.} {\bf 2009}, {\em 501},~941--947.
\newblock {\url{https://doi.org/10.1051/0004-6361/200811191}}.

\bibitem[{Flower}(1996)]{1996ApJ...469..355F}
{Flower}, P.J.
\newblock {Transformations from Theoretical Hertzsprung-Russell Diagrams to
  Color-Magnitude Diagrams: Effective Temperatures, B-V Colors, and Bolometric
  Corrections}.
\newblock {\em Astrophys. J. } {\bf 1996}, {\em 469},~355.
\newblock {\url{https://doi.org/10.1086/177785}}.

\bibitem[{Torres}(2010)]{2010AJ....140.1158T}
{Torres}, G.
\newblock {On the Use of Empirical Bolometric Corrections for Stars}.
\newblock {\em Astron. J.} {\bf 2010}, {\em 140},~1158--1162.
\newblock {\url{https://doi.org/10.1088/0004-6256/140/5/1158}}.

\bibitem[{Wang} and {Zhong}(2018)]{2018A&A...619L...1W}
{Wang}, J.; {Zhong}, Z.
\newblock {Revisiting the mass-luminosity relation with an effective
  temperature modifier}.
\newblock {\em Astron. Astrophys.} {\bf 2018}, {\em 619},~L1.
\newblock {\url{https://doi.org/10.1051/0004-6361/201834109}}.

\bibitem[{Griffiths} et~al.(1988){Griffiths}, {Hicks}, and
  {Milone}]{1988JRASC..82....1G}
{Griffiths}, S.C.; {Hicks}, R.B.; {Milone}, E.F.
\newblock {A re-examination of mass-luminosity relations from binary-star
  data}.
\newblock {\em J. R. Astron. Soc. Can.} {\bf 1988}, {\em 82},~1--12.

\bibitem[{Demircan} and {Kahraman}(1991)]{1991ApSS.181..313D}
{Demircan}, O.; {Kahraman}, G.
\newblock {Stellar Mass / Luminosity and Mass / Radius Relations}.
\newblock {\em Astrophys. Space Sci.} {\bf 1991}, {\em 181},~313--322.
\newblock {\url{https://doi.org/10.1007/BF00639097}}.

\bibitem[{Eker} et~al.(2015){Eker}, {Soydugan}, {Soydugan}, {Bilir}, {Yaz
  G{\"o}k{\c{c}}e}, {Steer}, {T{\"u}ys{\"u}z}, {{\c{S}}eny{\"u}z}, and
  {Demircan}]{2015AJ....149..131E}
{Eker}, Z.; {Soydugan}, F.; {Soydugan}, E.; {Bilir}, S.; {Yaz G{\"o}k{\c{c}}e},
  E.; {Steer}, I.; {T{\"u}ys{\"u}z}, M.; {{\c{S}}eny{\"u}z}, T.; {Demircan}, O.
\newblock {Main-Sequence Effective Temperatures from a Revised Mass-Luminosity
  Relation Based on Accurate Properties}.
\newblock {\em Astron. J.} {\bf 2015}, {\em 149},~131.
\newblock {\url{https://doi.org/10.1088/0004-6256/149/4/131}}.

\bibitem[{Moya} et~al.(2018){Moya}, {Zuccarino}, {Chaplin}, and
  {Davies}]{2018ApJS..237...21M}
{Moya}, A.; {Zuccarino}, F.; {Chaplin}, W.J.; {Davies}, G.R.
\newblock {Empirical Relations for the Accurate Estimation of Stellar Masses
  and Radii}.
\newblock {\em Astrophys. J. Suppl. Ser.} {\bf 2018}, {\em 237},~21.
\newblock {\url{https://doi.org/10.3847/1538-4365/aacdae}}.

\bibitem[{Malkov} et~al.(2010){Malkov}, {Sichevskij}, and
  {Kovaleva}]{2010MNRAS.401..695M}
{Malkov}, O.Y.; {Sichevskij}, S.G.; {Kovaleva}, D.A.
\newblock {Parametrization of single and binary stars}.
\newblock {\em Mon. Not. R. Astron. Soc.} {\bf 2010}, {\em 401},~695--704.
\newblock {\url{https://doi.org/10.1111/j.1365-2966.2009.15696.x}}.

\bibitem[{Bressan} et~al.(2012){Bressan}, {Marigo}, {Girardi}, {Salasnich},
  {Dal Cero}, {Rubele}, and {Nanni}]{2012MNRAS.427..127B}
{Bressan}, A.; {Marigo}, P.; {Girardi}, L.; {Salasnich}, B.; {Dal Cero}, C.;
  {Rubele}, S.; {Nanni}, A.
\newblock {PARSEC: Stellar tracks and isochrones with the PAdova and TRieste
  Stellar Evolution Code}.
\newblock {\em Mon. Not. R. Astron. Soc.} {\bf 2012}, {\em 427},~127--145.
\newblock {\url{https://doi.org/10.1111/j.1365-2966.2012.21948.x}}.

\bibitem[{Eker} et~al.(2018){Eker}, {Bak{\i}{\c{s}}}, {Bilir}, {Soydugan},
  {Steer}, {Soydugan}, {Bak{\i}{\c{s}}}, {Ali{\c{c}}avu{\c{s}}}, {Aslan}, and
  {Alpsoy}]{2018MNRAS.479.5491E}
{Eker}, Z.; {Bak{\i}{\c{s}}}, V.; {Bilir}, S.; {Soydugan}, F.; {Steer}, I.;
  {Soydugan}, E.; {Bak{\i}{\c{s}}}, H.; {Ali{\c{c}}avu{\c{s}}}, F.; {Aslan},
  G.; {Alpsoy}, M.
\newblock {Interrelated main-sequence mass-luminosity, mass-radius, and
  mass-effective temperature relations}.
\newblock {\em Mon. Not. R. Astron. Soc.} {\bf 2018}, {\em 479},~5491--5511.
\newblock {\url{https://doi.org/10.1093/mnras/sty1834}}.

\bibitem[{Hunt} and {Reffert}(2023)]{2023A&A...673A.114H}
{Hunt}, E.L.; {Reffert}, S.
\newblock {Improving the open cluster census. II. An all-sky cluster catalogue
  with Gaia DR3}.
\newblock {\em Astron. Astrophys.} {\bf 2023}, {\em 673},~A114.
\newblock {\url{https://doi.org/10.1051/0004-6361/202346285}}.

\end{thebibliography}

\PublishersNote{}

\end{adjustwidth}
\end{document}